\documentclass[letterpaper,aps,prd,twocolumn,showpacs,superscriptaddress,groupedaddress]{revtex4}  

\usepackage{graphicx}  
\usepackage{dcolumn}   
\usepackage{bm}        
\usepackage{amssymb}   
\usepackage{amsmath}
\graphicspath{{figures/}{figures/}}
\usepackage{atlasphysics_modified}
\usepackage{subfigure}
\usepackage{rotating}
\usepackage{mathrsfs}
\usepackage[hyperindex,breaklinks]{hyperref} 

\usepackage{preprintcover}  
\PreprintCoverPaperTitle{Measurement of Upsilon production in\\ 7~TeV {\boldmath $pp$} collisions at ATLAS}  
\PreprintIdNumber{CERN-PH-EP-2012-295}  
\PreprintCoverAbstract{  
  Using 1.8~fb$^{-1}$ of \pp\ collisions at a center-of-mass energy of 7 \TeV\ recorded by
  the ATLAS detector at the Large Hadron Collider, we present measurements of the production
  cross sections of $\varUpsilon(1\mathrm{S},2\mathrm{S},3\mathrm{S})$ mesons.
  $\varUpsilon$ mesons are reconstructed using the di-muon decay mode.
  Total production cross sections for $p_{\rm T}<$70~GeV and in the rapidity interval $\absyUps<$2.25 are measured to be
  8.01$\pm$0.02$\pm$0.36$\pm$0.31~nb, 2.05$\pm$0.01$\pm$0.12$\pm$0.08~nb, 0.92$\pm$0.01$\pm$0.07$\pm$0.04~nb,
  respectively, with uncertainties separated into statistical, systematic, and luminosity measurement effects.
  In addition, differential cross section times di-muon branching fractions for \UpsXS{1}, \UpsXS{2}, and \UpsXS{3}
  as a function of $\varUpsilon$ transverse momentum $p_{\rm T}$ and rapidity are presented.
  These cross sections are obtained assuming unpolarized production.
  If the production polarization is fully transverse or longitudinal with no azimuthal dependence in the helicity frame
  the cross section may vary by approximately $\pm$20$\%$. If a non-trivial azimuthal dependence is
  considered, integrated cross sections may be significantly enhanced by a factor of two or more.
  We compare our results to several theoretical models of $\varUpsilon$ meson production,
  finding that none provide an accurate description of our data over the full range of $\varUpsilon$
  transverse momenta accessible with this dataset.
}  
\PreprintJournalName{PRD}  

\begin{document}

\title{Measurement of Upsilon production in 7~TeV {\boldmath $pp$} collisions at ATLAS}
\author{ATLAS Collaboration}

\begin{abstract}
  Using \intLumi\ of \pp\ collisions at a center-of-mass energy of 7 \TeV\ recorded by
  the ATLAS detector at the Large Hadron Collider, we present measurements of the production
  cross sections of $\varUpsilon(1\mathrm{S},2\mathrm{S},3\mathrm{S})$ mesons.
  $\varUpsilon$ mesons are reconstructed using the di-muon decay mode.
  Total production cross sections for $p_{\rm T}< 70$~GeV and in the rapidity interval $\absyUps<2.25$ are measured to be
  \totXSoneS , \totXStwoS , and \totXSthreeS ,
  respectively, with uncertainties separated into statistical, systematic, and
  luminosity measurement effects.
  In addition, differential cross section times di-muon branching fractions for \UpsXS{1}, \UpsXS{2}, and \UpsXS{3}
  as a function of $\varUpsilon$ transverse momentum $p_{\rm T}$ and rapidity are presented.
  These cross sections are obtained assuming unpolarized production.
  If the production polarization is fully transverse or longitudinal with no azimuthal dependence in the helicity frame
  the cross section may vary by approximately $\pm 20\%$. If a non-trivial azimuthal dependence is
  considered, integrated cross sections may be significantly enhanced by a factor of two or more.
  We compare our results to several theoretical models of $\varUpsilon$ meson production,
  finding that none provide an accurate description of our data over the full range of $\varUpsilon$
  transverse momenta accessible with this dataset.
\end{abstract}

\maketitle


\hyphenation{off-line}
        
\newpage

\section{Introduction}
\label{sect:Introduction}
Since the discovery of the \Jpsi\ and \Ups\ mesons~\cite{Jpsi:discovery,Upsilon:discovery}
the study of heavy quark-antiquark systems has provided valuable input for our understanding of Quantum Chromodynamics (QCD). 
However, despite being one of the simplest systems in QCD it has proven difficult historically to describe the production properties of these states adequately.
Results on \Jpsi\ and \Ups\ hadroproduction and 
polarization\,\cite{previous1,CDF:upsilon,previous2,CMS:upsilon,ATLAS:jpsi,ATLAS:upsilon,ALICE:jpsi,LHCb:jpsi,LHCb:upsilon,CMS:upsilonpol}
exhibit inconsistencies between measurements and theoretical predictions\,\cite{Brambilla:2011}.

Measurements at the Large Hadron Collider (LHC) of differential production spectra
of various charmonium and bottomonium states together with measurement of their
spin-alignments, prompt double-quarkonia production, and
production of quarkonia in association with photons, vector bosons or open heavy-flavor final 
states will allow discrimination between
different theoretical approaches based on color-singlet corrections\,\cite{CSM,NNLOUpsilon,NNLOUpsilon2},
color-octet terms\,\cite{COM}, the $k_{\rm T}$-factorization approach\,\cite{KTFAC} and other production models\,\cite{CEM},
and provide additional input toward an improved understanding of quarkonium hadroproduction.

Studies of bottomonium production complement concurrent studies of charmonium systems
due to the larger mass of the bottom quark compared to the charm quark,
allowing more dependable theoretical calculations than in the charmonium family, which suffer from poor perturbative convergence\,\cite{COM}.
Extension of cross-section measurements to higher meson transverse momenta provides valuable input to improvements in the theoretical description since in this regime 
different processes can dominate and, experimentally, the impact of spin-alignment uncertainties are mitigated.

Taking advantage of the large increase in integrated luminosity delivered by the LHC in 2011, 
this paper updates a previous measurement\,\cite{ATLAS:upsilon} and reports the \UpsXS{1} cross section 
presented differentially in two intervals of absolute $\varUpsilon$ rapidity\,\cite{ATLASdetectorfootnote},
$|y^\varUpsilon|$, and fifty intervals of $\varUpsilon$ transverse momentum, \pTUps,
and as a $p_{\rm T}$-integrated result in forty-five bins of absolute $\varUpsilon$ rapidity. 
In addition, new measurements of the equivalent \UpsXS{2} and \UpsXS{3} differential spectra and their production ratios relative to the \UpsXS{1} are reported.

$\varUpsilon$ production can proceed {\em directly}, where the \Ups\ meson of interest is produced in the hard interaction, or via the production of an excited state
which subsequently decays. This so-called ``feed-down'' contribution complicates the theoretical description of
quarkonium production as calculation of color-singlet $P$-wave and higher orbital angular momentum quarkonium production suffers from the presence of infrared divergences\,\cite{Brambilla:2004wf}.
From the experimental perspective, separation of direct and feed-down contributions is hindered by the small mass splitting between the bottomonium states, which impedes the 
detection of additional decay products that would indicate indirect production.
Contributions from feed-down vary between the \UpsXS{1}, \UpsXS{2} and \UpsXS{3} states due to the changing presence of various kinematically-allowed decays 
and influence the inclusive production rate.
Study of the \Ups\ production ratios as a function of kinematic variables as presented here thus provides an indirect but precise measure of these feed-down contributions.

This new analysis extends the \pT\ range of our previous cross-section result\,\cite{ATLAS:upsilon} to 70\,\GeV, 
at which new contributions to $\varUpsilon$ production\,\cite{NNLOUpsilon,NNLOUpsilon2} such as from associated $\varUpsilon+b\overline{b}$ production
may play a more important role and the impact of 
the dependence of the production cross section on the spin-alignment of the $\varUpsilon$ is relatively small. 
Both the fiducial cross section, 
measured in the kinematic region with muon transverse momentum $p_{\rm T}^\mu>4$\,\GeV\ and muon pseudorapidity $|\eta^\mu|<2.3$,
and the corrected cross section, which is defined in this paper as the cross section in the $p_{\rm T}$--$\eta$ phase space of 
the $\varUpsilon$ corrected for the acceptance of the decay products to the
full \Ups\ production phase space, are reported. In the former case, the results have no dependence on assumptions about \Ups\ spin
alignment, while the latter measurements are more easily compared to model predictions and to the results of other experiments.

\section{The ATLAS Detector}
\label{sect:Detector}
The ATLAS detector\,\cite{ATLAS:det} is composed of an inner tracking
system, calorimeters, and a muon spectrometer.
The inner detector directly surrounds the interaction point and
consists of a silicon pixel detector, a silicon microstrip detector, and a
transition radiation tracker, all immersed in a 2\,T axial magnetic
field. It covers the pseudorapidity\,\cite{ATLASdetectorfootnote} range $\abseta<2.5$ and is
enclosed by a calorimeter system containing electromagnetic and
hadronic sections. Surrounding the calorimeters is the large muon
spectrometer built with three air-core toroids. This spectrometer is
equipped with precision detectors (monitored drift tubes and cathode
strip chambers) that provide precise measurements in the bending plane
within the pseudorapidity range $\abseta<2.7$. In addition,
resistive plate and thin gap chambers with fast response times are
used primarily to construct muon triggers in the ranges $\abseta<1.05$ and $1.05<\abseta<2.4$ respectively,
but are also used to provide position measurements in the non-bending
plane and to improve pattern recognition and track reconstruction.
Momentum measurements in the muon spectrometer are based on track
segments formed in at least two of the three precision chamber planes.

The ATLAS detector employs a three-level trigger system\,\cite{ATLAS:trig}, which reduces the 20\,MHz proton bunch collision rate
to the several hundred Hz transfer rate to mass storage.
The Level-1 muon trigger searches for hit coincidences between
different muon trigger detector layers inside pre-programmed
geometrical windows that bound the path of triggered muons of given transverse
momentum and provide a rough estimate of its position within the
pseudorapidity range $\abseta<2.4$. 
At Level-1, muon candidates are reported in ``regions of interest'' (RoIs). 
Only a single muon can be associated with a given RoI of spatial extent 
$\dphi \times \deta \approx 0.1 \times 0.1$. This limitation has a small 
effect on the trigger efficiency for \Ups\ mesons, corrected for in the analysis using a data-driven method based on analysis of $J/\psi\to\mu^+\mu^-$ and $\varUpsilon\to\mu^+\mu^-$ decays.
The Level-1 trigger is followed in sequence by two subsequent higher-level, software-based trigger selection
stages. Muon candidates reconstructed at these higher levels incorporate, with increasing precision, information from both
the muon spectrometer and the inner detector and reach position and momentum resolution close to that provided by the offline muon reconstruction.

\section{Dataset and Event Selection}
\label{sect:Data}
Data for this study were collected during the 2011 LHC
proton-proton running period between March and August
using a trigger that requires the presence of two muon candidates with
opposite charges that 
are subject to a fit constraining them to a common vertex while taking into account
track parameter uncertainties. A very loose selection on vertex $\chi^2$ which is fully efficient for signal candidates, was imposed
to ensure proper fit convergence, as well as the   
requirement of opposite charge and that $p^\mu_{\rm T}>4$\,\GeV\ and $|\eta^\mu|<2.3$.
This trigger was largely unprescaled and collected data at a rate of approximately 20~Hz in this period of data-taking.

Events are required to contain at least one 
primary vertex candidate that has at least five tracks with $p_{\rm T}>0.4$\,\GeV,
and at least two muons identified by
associating candidates found in the muon spectrometer
with tracks reconstructed in the inner
detector\,\cite{ATLAS:upsilon,ATLAS:det}.
Multiple scattering in the calorimeters and toroids of the ATLAS detector degrades the muon
spectrometer resolution for low energy particles. 
As the majority of muons selected for this analysis have low momentum,
we assign values for parameters such as \pT\ and $\eta$ to the muons based on
track fits using inner detector information only. 
To ensure accurate inner detector measurements, each muon track must contain at
least six silicon microstrip detector hits and at least one pixel detector hit. Muon candidates passing
these criteria are required to have $\pTmu>4$\,\GeV\ and $\etamu<2.3$ and a successful fit to a common vertex.

Good spatial matching ($\dR \equiv \sqrt{(\dphi)^2 + (\deta)^2} < 0.01$) between the muon candidate in both the offline reconstruction and the trigger 
used to select the event is required to facilitate data-driven estimates of the di-muon trigger efficiency.
Furthermore, both muons forming an \Ups\ candidate must be associated to their trigger-level candidates in this manner.
In this way, the efficiency of the trigger 
requirements on the di-muon candidate are incorporated into the trigger efficiency correction.
All di-muon candidates passing these criteria are retained for analysis.

\begin{figure*}[htbp]
\begin{center}
  \includegraphics[width=0.49\textwidth]{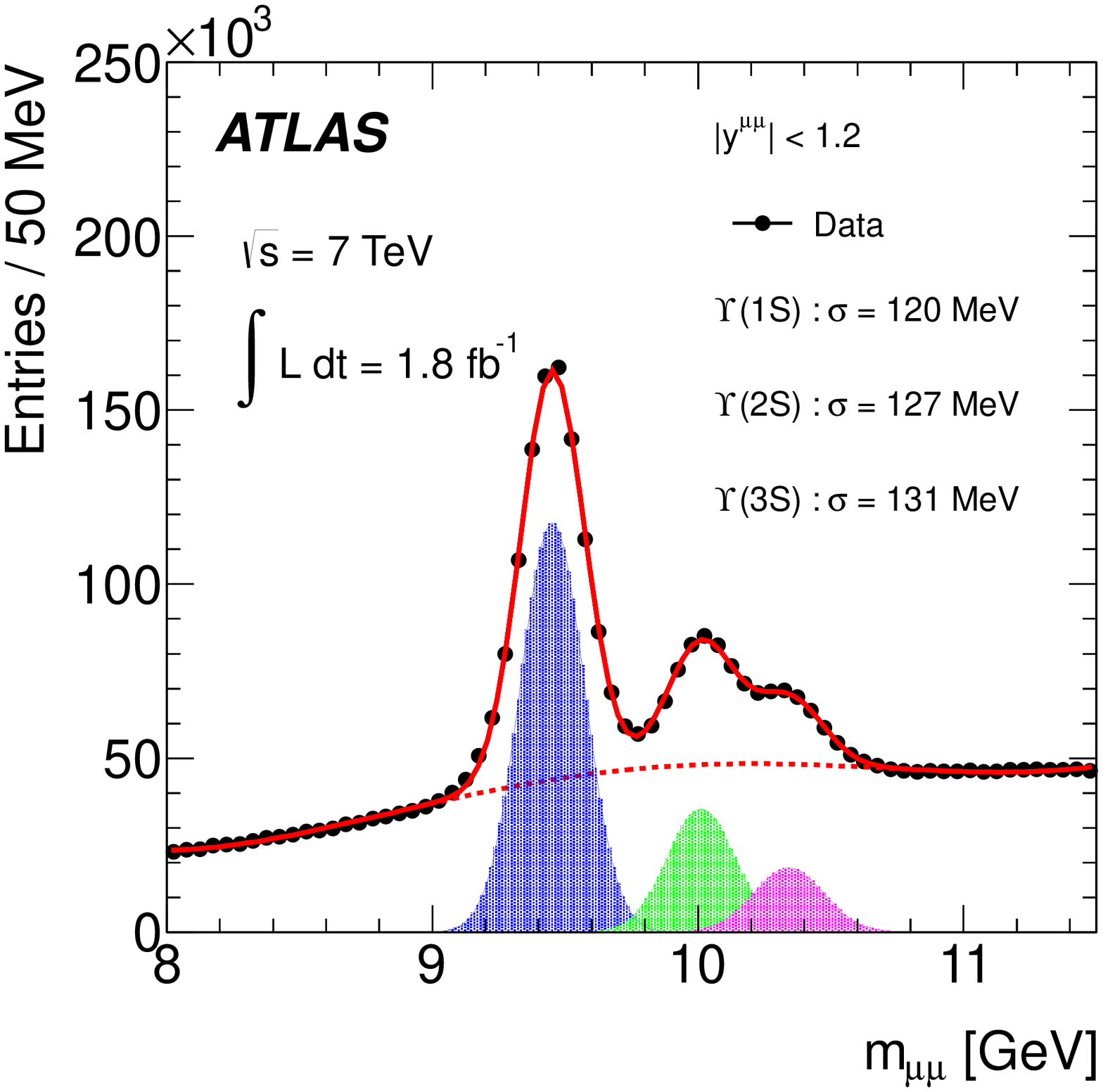}
  \includegraphics[width=0.49\textwidth]{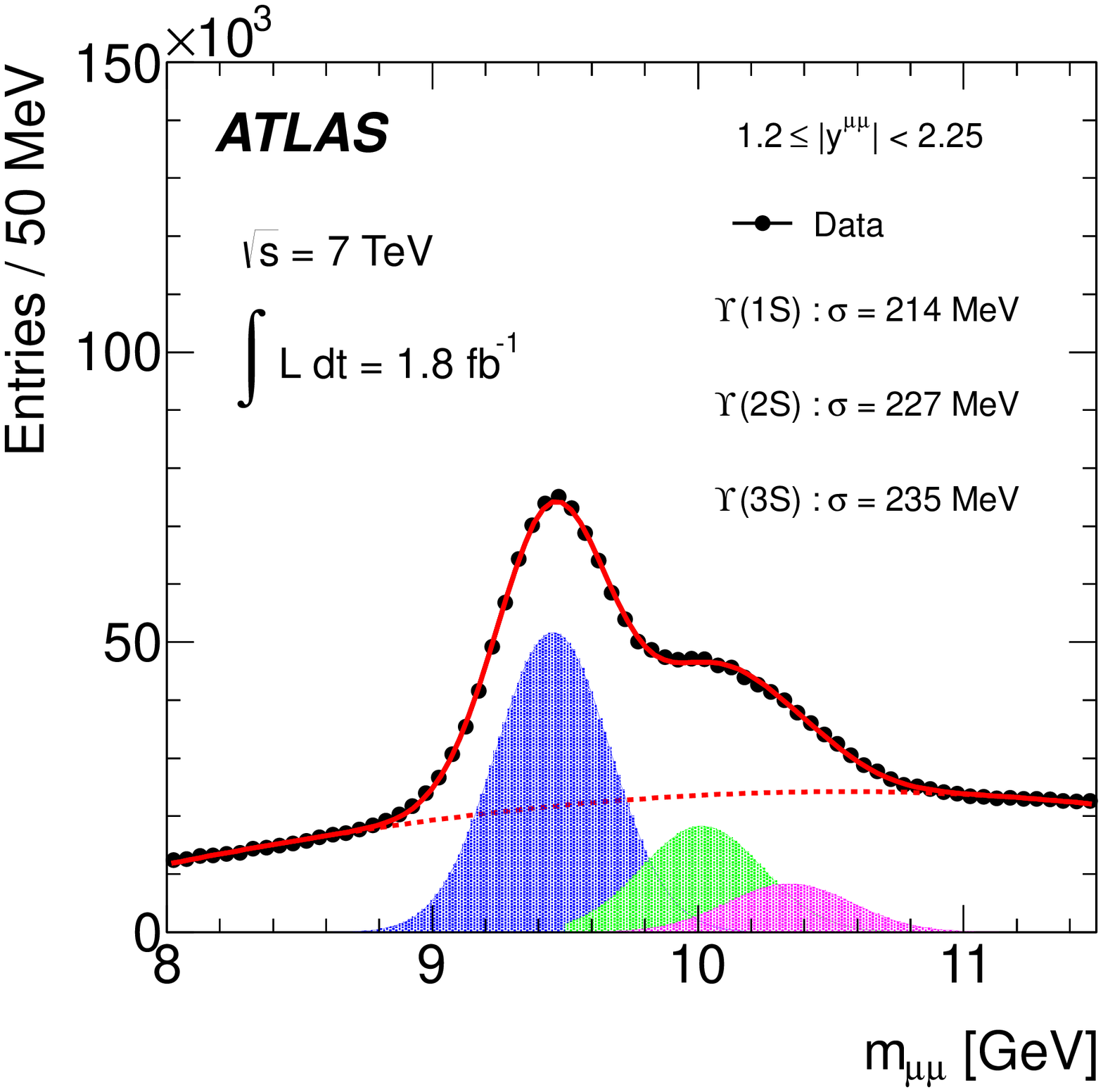}
  \caption{The di-muon invariant mass spectum for events used in this
    analysis. Separate spectra are shown for those events with the
    di-muon candidate (left) in the central region of the detector ($\absymumu<1.2$)
    and (right) in the forward region ($1.2\leq\absymumu<2.25$). 
    Overlaid are individual shapes of the fitted $\varUpsilon(n\mathrm{S})$ signals (shaded regions), background-only fit (dashed curve)
    and the total signal plus background shape (solid curve). The background shape is modelled here by a fourth-order polynomial and the signal peaks each
    modelled by a single Gaussian. Also quoted are the fitted mass resolutions of each of the three signal peaks, determined from the fit with a common resolution parameter 
    scaling with invariant mass.}
  \label{fig:Mmumu}
\end{center}
\end{figure*}

The distribution of the invariant mass of the \mumu\ system is shown 
in Fig.~\ref{fig:Mmumu} for the selected di-muon candidates.
As is apparent from the plot, the mass resolution (120\,MeV) for di-muon candidates
detected in the central region of the detector ($\absymumu<1.2$) is significantly
better than the resolution (214\,MeV) for those candidates falling into the forward region 
($1.2\leq\absymumu<2.25$). A total of \scinot{3.9}{6} and \scinot{2.3}{6} candidates 
with $8<\Mmumu<11.5$\,\GeV\ are reconstructed in the central and forward regions respectively.

\section{Cross Section Determination}
\label{sect:Xsect}
Differential \Ups\ cross sections are measured according to the relation:
\begin{equation}
  \frac{d^2 \sigma}{d\pT dy} \times
  \mathrm{Br}({\UpsMuMu}) = 
  \frac{N_{\Ups}}{\int \mathcal{L} dt \times \Delta \pT \times \Delta |y|}
\label{eqn:xsect}
\end{equation}
where  $\mathrm{Br}({\UpsMuMu})$ represents the appropriate branching fraction of the \UpsnS\ to di-muons, 
$\int \mathcal{L} dt$ is the integrated luminosity,
$\Delta \pT$ and $\Delta |y|$ are the bin sizes in \Ups\ transverse
momentum and rapidity, respectively,
and $N_{\Ups}$ is the corrected number of observed $\varUpsilon(1\mathrm{S})$, $\varUpsilon(2\mathrm{S})$, or $\varUpsilon(3\mathrm{S})$ mesons
in a bin. Corrections are applied to account for selection efficiencies, bin migration effects due to finite detector resolution, 
and, in the case of corrected cross-section measurements, acceptance.  

Determination of the cross sections proceeds through several
steps. Firstly, a weight is determined for each selected di-muon candidate
equal to the inverse of the total efficiency for the candidate.
Secondly, a fit is performed to the distribution of weighted events binned
in di-muon invariant mass, \Mmumu, to determine the number of
\UpsnS\ mesons (with $n$ = 1,2,3) produced in each 
(\pTmumu, \ymumu) bin.
Thirdly, these values are corrected for (\pTmumu, \ymumu) 
bin migrations. 
Finally, the differential cross section multiplied by the \UpsMuMu\ branching
fraction is calculated for each state using the integrated
luminosity and the \pT\ and $y$ bin widths as in
Eq.~(\ref{eqn:xsect}). 

The weight, $w$, for each \Ups\ candidate includes the fraction of
produced \UpsMuMu\ decays with both muons falling
into the kinematic region $\pTmu>4$\,\GeV\ and $\etamu<2.3$
(the acceptance, \acc , is used only in calculating corrected cross sections),
the probability that a candidate falling within the acceptance passes
the offline reconstruction requirements
(the reconstruction efficiency, \effReco ),
and the probability that a reconstructed event passes the trigger
selection (the trigger efficiency, \effTrig ). 
The weights assigned to a given candidate when calculating the
fiducial (\wFid ) and corrected (\wTot ) cross sections are then given by:
\begin{align}
\begin{split}
  \wFid &= (\effReco \cdot \effTrig )^{-1},\\
  \wTot &= (\acc \cdot \effReco \cdot \effTrig )^{-1}.
\end{split}
\label{eqn:weight}
\end{align}

\subsection{Acceptance} 
The kinematic acceptance ${\cal {A}}(p_{\rm T},y)$ is the probability that the muons
from an $\varUpsilon$ with transverse momentum $p_{\rm T}$ and rapidity $y$
fall into the fiducial volume of the detector defined by the $\pTmu>4$\,\GeV\ and $\etamu<2.3$
selection applied to each muon in the di-muon pair.
In order to calculate the acceptance as a function of the rapidity and transverse momentum of each of the \Ups\ states, 
taking into account possible angular dependences in their decays,
we use an analytic formula describing the decay of \Ups\ states in their decay frame\,\cite{faccioli},
\begin{align}
\begin{split}
  \frac{d^2 N}{d\cos \theta^* d\phi^*} \propto 1 &+ \lambda_{\theta} \cos^2 \theta^* + \lambda_{\phi} \sin^2 \theta^* \cos 2\phi^* \\
&+ \lambda_{\theta \phi} \sin 2\theta^* \cos \phi^*.
\label{eqn:spin}
\end{split}
\end{align}

The helicity frame is used, where $\theta^*$ is the polar angle between the $\mu^+$ momentum in the \Ups\ decay frame and the direction
of the \Ups\ momentum in the laboratory frame. The corresponding azimuthal angle $\phi^*$ is the angle between the quarkonium production 
plane (defined in the quarkonium decay frame by the two momenta of the incoming protons in that frame) and the
quarkonium decay plane in the lab frame.

The $\lambda_i$ coefficients parameterize the spin-alignment state of the \Ups.
In some parts of the phase space the acceptance may depend on these parameters quite strongly. 
We have identified five extreme cases that lead to the largest possible variations of acceptance
within the phase space of this measurement and define an envelope in which the results may vary under any physically-allowed spin-alignment assumption:
isotropic distribution independent of $\theta^*$ and $\phi^*$
($\lambda_{\theta} = \lambda_{\phi} = \lambda_{\theta \phi} = 0$, labeled FLAT);
longitudinal alignment
($\lambda_{\theta} = -1, \lambda_{\phi} = \lambda_{\theta \phi} = 0$, labeled LONG);
and three types of transverse alignment
($\lambda_{\theta} = +1, \lambda_{\phi} = \lambda_{\theta \phi} = 0$, labeled $T\mathrm{+0}$;
$\lambda_{\theta} = +1, \lambda_{\phi} = +1, \lambda_{\theta \phi} = 0$, labeled $T\mathrm{++}$;
$\lambda_{\theta} = +1, \lambda_{\phi} = -1, \lambda_{\theta \phi} = 0$, labeled $T\mathrm{+-}$).
The central values of our measurements are derived under the isotropic production assumption. 
The spread in corrected cross-section results derived under all five assumptions is used to quantify the full envelope of 
possible variations of the result due to spin-alignment.
Some constraints on physically-allowed $\lambda_i$ combinations exist\,\cite{faccioli,palestini}. In particular, $\lambda_{\theta \phi}$ must be zero
for $\lambda_{\theta}=\pm1$ and any $\lambda_{\phi}$, and $|\lambda_{\theta \phi}|\leq 0.5$ for $\lambda_{\theta} = \lambda_{\phi} = 0$.
Acceptance corrections with non-zero $\lambda_{\theta \phi}$ were found to lead to corrections with smaller or equal variations from the isotropic scenario
than any parameter-space scenarios on the $\lambda_{\theta}$--$\lambda_{\phi}$ plane. Acceptance weights for each polarization scenario,
including those with non-zero $\lambda_{\theta \phi}$ can be found in \textsc{hepdata}\,\cite{HEPDATA}.

\begin{figure}[h!tb]
  \begin{center}
    \includegraphics[width=0.49\textwidth]{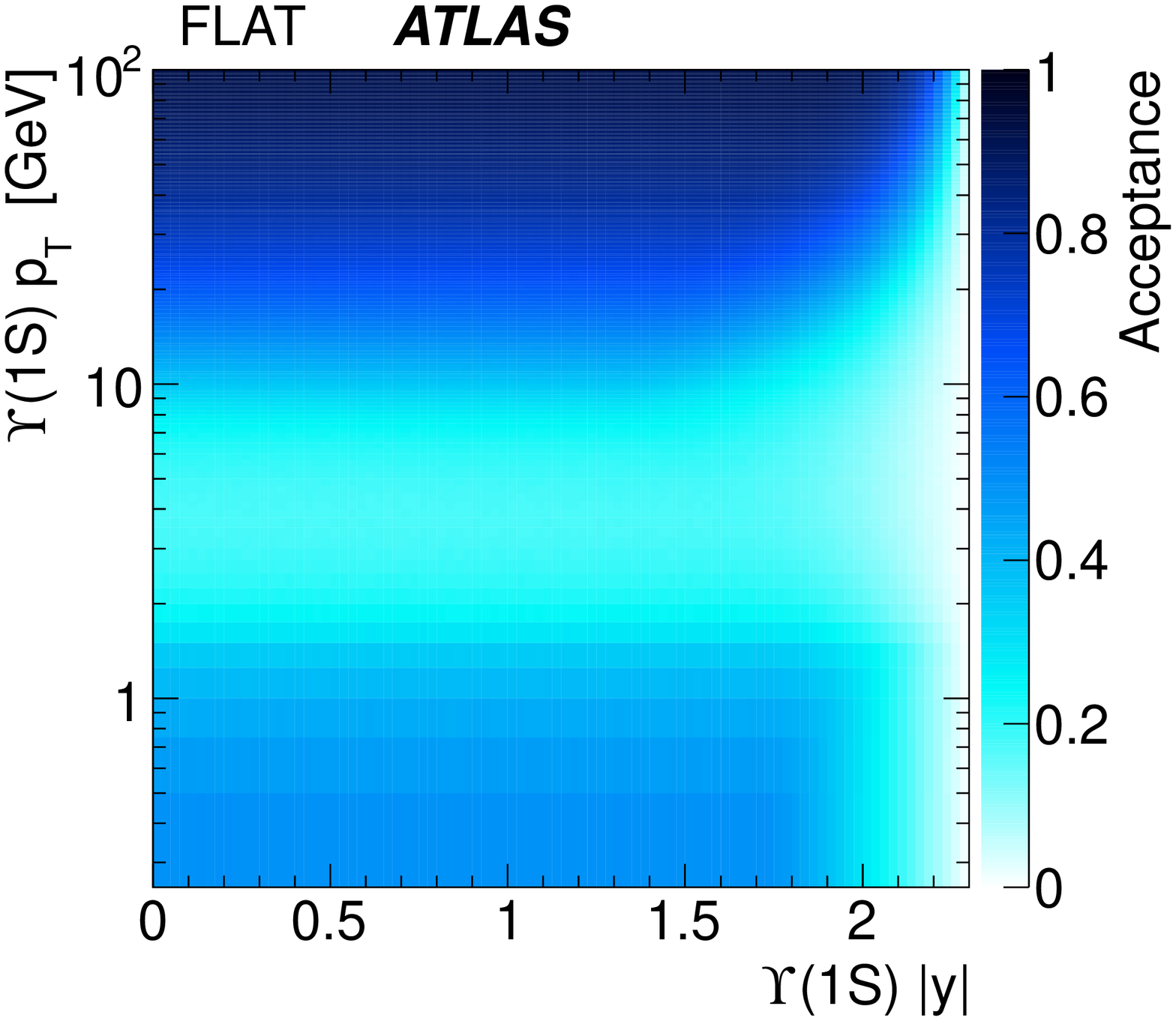}
    \includegraphics[width=0.49\textwidth]{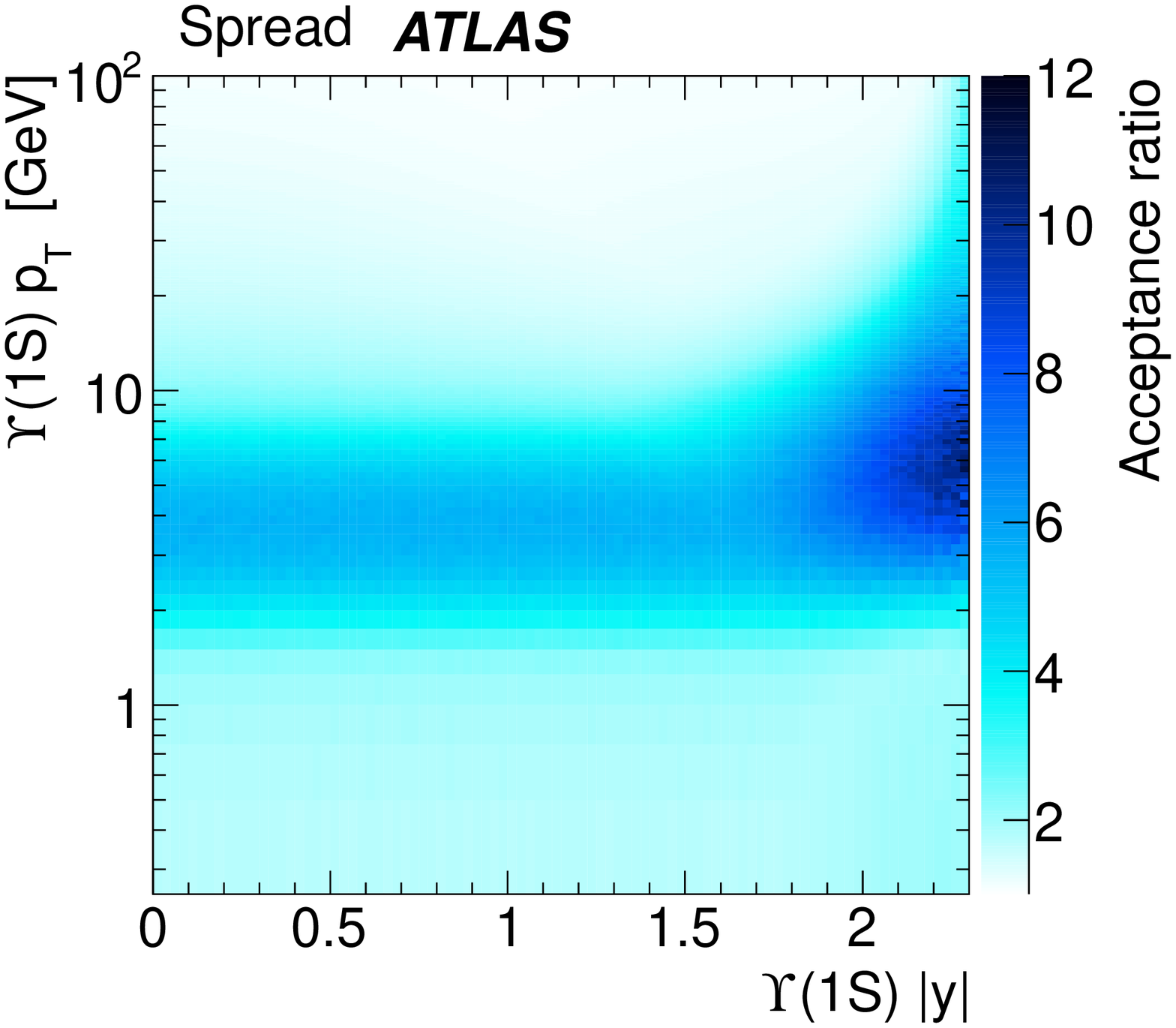}
    \caption{(Top) Acceptance as a function of $\varUpsilon$ $p_{\rm T}$ and rapidity for the \UpsXS{1} in the unpolarized acceptance scenario; 
      (bottom) the ratio of largest to smallest acceptance correction between extreme spin-alignment scenarios.}
    \label{fig:Acceptance}
  \end{center}
\end{figure}

Figure~\ref{fig:Acceptance} presents the two-dimensional acceptance map for $\varUpsilon(1\mathrm{S})$, as a function of the $\varUpsilon$ transverse momentum and absolute rapidity
in the unpolarized (FLAT) acceptance scenario as well as the maximum relative variation (ratio of largest to smallest acceptance correction) 
of the acceptance maps across all spin-alignment scenarios.

Acceptance is significantly decreased at very high rapidities due to the muon pseudorapidity requirement of $\etamu<2.3$.
When $\varUpsilon$ are produced at rest, both muons are likely to be reconstructed in the muon spectrometer on opposite sides of the detector. When the $\varUpsilon$ has
a sufficient boost, the two muons are more likely to be found in the same side of the detector and the probability that at least one muon is below the required \ptmu\ threshold increases,
leading to a drop in acceptance. At higher $p_{\rm T}$ the $\varUpsilon$ imparts greater momentum to the two muons, the likelihood of a muon being below threshold decreases
and the acceptance asymptotically approaches 100\%. As $\varUpsilon$ transverse momentum increases the variation of the acceptance corrections between spin-alignment scenarios decreases,
becoming $\lesssim 10\%$ at the largest $p_{\rm T}$ ranges studied.
Similar behaviours are observed in the acceptance maps for the $\varUpsilon(2\mathrm{S})$ and $\varUpsilon(3\mathrm{S})$.

This spin-alignment systematic uncertainty only applies to the corrected cross-section measurements. For fiducial cross sections, 
measured in the kinematic region $\pTmu>4$\,\GeV\ and $\etamu<2.3$, we do not have to correct our results to the full phase space of
produced $\varUpsilon$ mesons. Spin-alignment can thus affect only our estimates of the reconstruction and trigger efficiencies. Since we
measure these in many bins of muon \pT\ and $\eta$, effects from differing distributions of events over individual bins due to
different spin-alignment assumptions are negligible compared to other sources of systematic uncertainty.

\subsection{Reconstruction and trigger efficiencies}
The efficiency of the offline reconstruction criteria
for events with muons within the fiducial region is given by:
\begin{align}
\begin{split}
  \effReco = 
  & \effTrk (p_{\rm T1},\eta_{1}) \times \effTrk(p_{\rm T2}, \eta_{2})\\
  & \times
  \effMu (p^\mu_{\rm T1},\, q_{1}\cdot \eta^\mu_{1}) \times
  \effMu (p^\mu_{\rm T2},\, q_{2}\cdot \eta^\mu_{2}).
\end{split}
\end{align}
In this equation, $q$ represents the charge of the muon and the identities of the two muons are labeled with indices $1,2$.
The efficiency of the track selection criteria, \effTrk, for tracks originating from real muons
is determined in Ref.\,\cite{ATLAS:jpsi} using data collected in 2010. 
Due to the presence of additional $pp$ interactions in the same and neighbouring bunch crossings in 2011 data the systematic uncertainty
on the tracking efficiency has increased, with \effTrk\ assessed to be $99\pm1.0\%$ over the whole kinematic range.

The efficiency to reconstruct a muon, \effMu, is derived using a tag-and-probe method on \JpsiMuMu\ data.
The {\it tag} muon corresponds to a muon candidate with $\pTmu>4$\,\GeV\ and $\etamu<2.4$, and must have fired a single-muon trigger in the event, 
as required by the trigger matching algorithms.
The {\it probe} track is only required to pass the inner detector track quality, \pT, and $\eta$ cuts and be consistent with having the same vertex as the identified tag muon.
This technique provides a sample of muon candidates unbiased with respect to both trigger and offline reconstruction, and with favorable signal to background ratio.

The muon reconstruction efficiency is then derived in two-dimensional $p_{\rm T}$--$\eta$ bins (fourteen muon \pT\ intervals and twenty-six charge-signed muon
pseudorapidity [\qeta] intervals). The ratio of the fitted \JpsiMuMu\ signal yield for those probe tracks identified as muons
to the fitted signal yield for all probe tracks in these double-differential intervals is identified as the single-muon reconstruction efficiency.

Due to the toroidal magnetic field, muons with positive (negative) charge are bent towards larger (smaller) pseudorapidity. This introduces a
charge dependence in the muon reconstruction efficiency. We calculate efficiencies as a function of charge-signed pseudorapidity as 
negative muons at positive rapidities are affected in the same manner as positive muons in negative rapidities. 
The charge dependence is particularly noticeable at very large $|\eta|$ where the muon can be bent outside of the geometrical acceptance of the detector. In particular at low $p_{\rm T}$,
where the muons of particular charge may be bent away from (rather than toward for the opposing charge) the middle/outer spectrometer planes, they will not be identified as muons.

The efficiency of the di-muon trigger to select events that have passed the offline selection criteria, \effTrig , is also calculated from
data. This can be factorized into three terms:
\begin{align}
\begin{split}
  \effTrig = 
  & \effRoI (p^\mu_{\rm T1},\, q_{1}\cdot \eta^\mu_{1}) \times
  \effRoI (p^\mu_{\rm T2},\, q_{2}\cdot \eta^\mu_{2}) \\
  & \times
  \cMuMu (\dR , \absymumu )
\label{eqn:effTrig}
\end{split}
\end{align}
where \effRoI\ is the efficiency of the trigger system to find an RoI 
for a single muon with transverse momentum, \pTmu, and charge-signed pseudorapidity, $q\cdot\etamu$,
and \cMuMu\ is a correction for effects related to the di-muon elements of the trigger.
This correction accounts for the di-muon vertex and opposite charge requirements, 
and for loss of efficiency in the di-muon trigger if the two muons are close enough together to register only as a single RoI.

The di-muon correction, \cMuMu, itself consists of two components
\begin{equation}
  \cMuMu (\dR, \ymumu ) = \cAsymp (\ymumu ) \times \cdR (\dR , \ymumu )
\label{eqn:cMuMu}
\end{equation}
each evaluated in three separate regions of di-muon rapidity:
barrel (\absymumu\ $\leq$ 1.0), barrel-endcap transition ($1.0<\absymumu\leq1.2$),
and endcap ($1.2\leq\absymumu<2.25$),
to account for the different behaviors of these corrections in these regions.
The correction \cAsymp\ is due to the effect of vertex and opposite charge requirements on the trigger and is determined from the efficiency at large di-muon angular separations.
No difference was observed from deriving the result as a function of $\Delta\eta$ and $\Delta\phi$ separately rather than versus $\Delta R$.
The asymptotic values are found using the ratio of candidate \JpsiMuMu\ decays selected
by the standard di-muon trigger to those selected by a similar di-muon trigger that makes no charge or vertex
requirements. These values are found to be $99.1\pm 0.4\%$, $97.5\pm 0.9\%$, $95.2\pm 0.4\%$ in the barrel,
transition, and endcap regions, respectively.

We extract this dependence on \dR\ in the same three regions of di-muon
rapidity as used for \cAsymp\ from a sample of offline reconstructed 
$J/\psi\to\mu^+\mu^-$ and $\varUpsilon\to\mu^+\mu^-$ candidates with $p^\mu_{\rm T2}>$8\,\GeV\
selected using a single-muon trigger with a threshold of 18\,\GeV. 
The 8\,\GeV\ requirement on the lower \pT\ muon is made to ensure that the efficiency for the trigger system to identify RoIs compatible with muons of $p_{\rm T}>4$\,\GeV\ 
reached its plateau value. The \dR\ dependence of \cdR\ is then extracted 
from the fraction of fitted di-muon candidates in a \dR\ interval within this control sample selected using auxiliary triggers 
that additionally pass the di-muon trigger used in this analysis.

The final component of the di-muon trigger efficiency, \effRoI, represents the single-muon 
trigger efficiency with a threshold of $\pTmu>4$\,\GeV. This is measured using well-reconstructed $J/\psi\to\mu\mu$
candidates in data that pass a single-muon trigger (with a threshold of 18\,\GeV).
The ratio of the yield of $J/\psi$ candidates (determined by fitting the invariant mass distributions) 
that pass both this single-muon trigger and the 4\,\GeV\ \ptmu\ threshold di-muon trigger used in this analysis,
to the yield of $J/\psi$ candidates that pass the single-muon trigger 
is identified as the single-muon trigger efficiency \effRoI\ in the \pTmu\ and $q\cdot\etamu$ interval considered.
In each case, the reconstructed muon(s) are matched to the muon(s) that triggered the event for each of the single or di-muon triggers.
The number of candidates passing the di-muon trigger is then further corrected by \cMuMu\ for di-muon correlation 
effects\,\cite{correlationfootnote}.

\begin{figure}[h!tbp]
\begin{center}
    \includegraphics[width=0.42\textwidth]{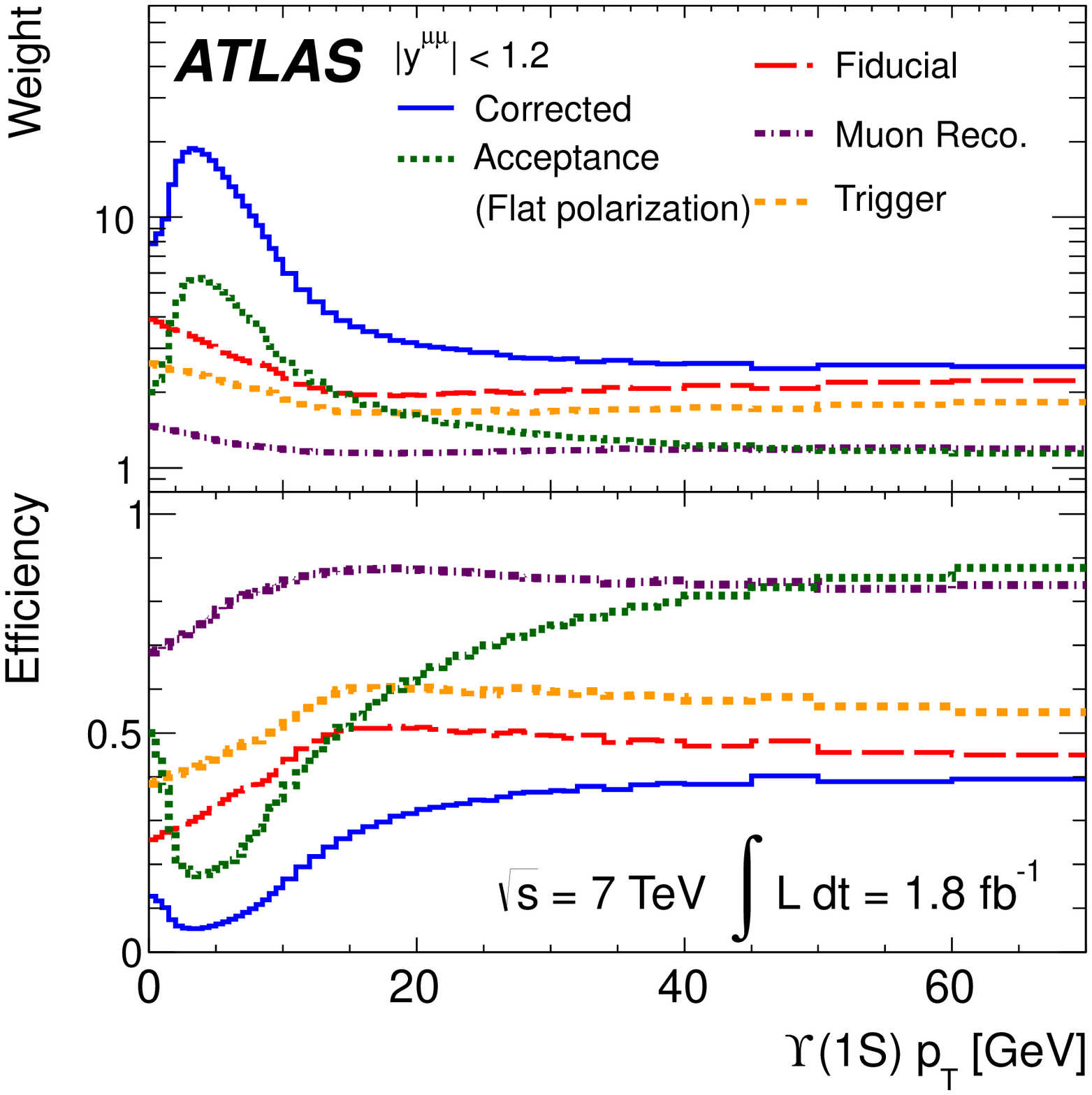}
    \includegraphics[width=0.42\textwidth]{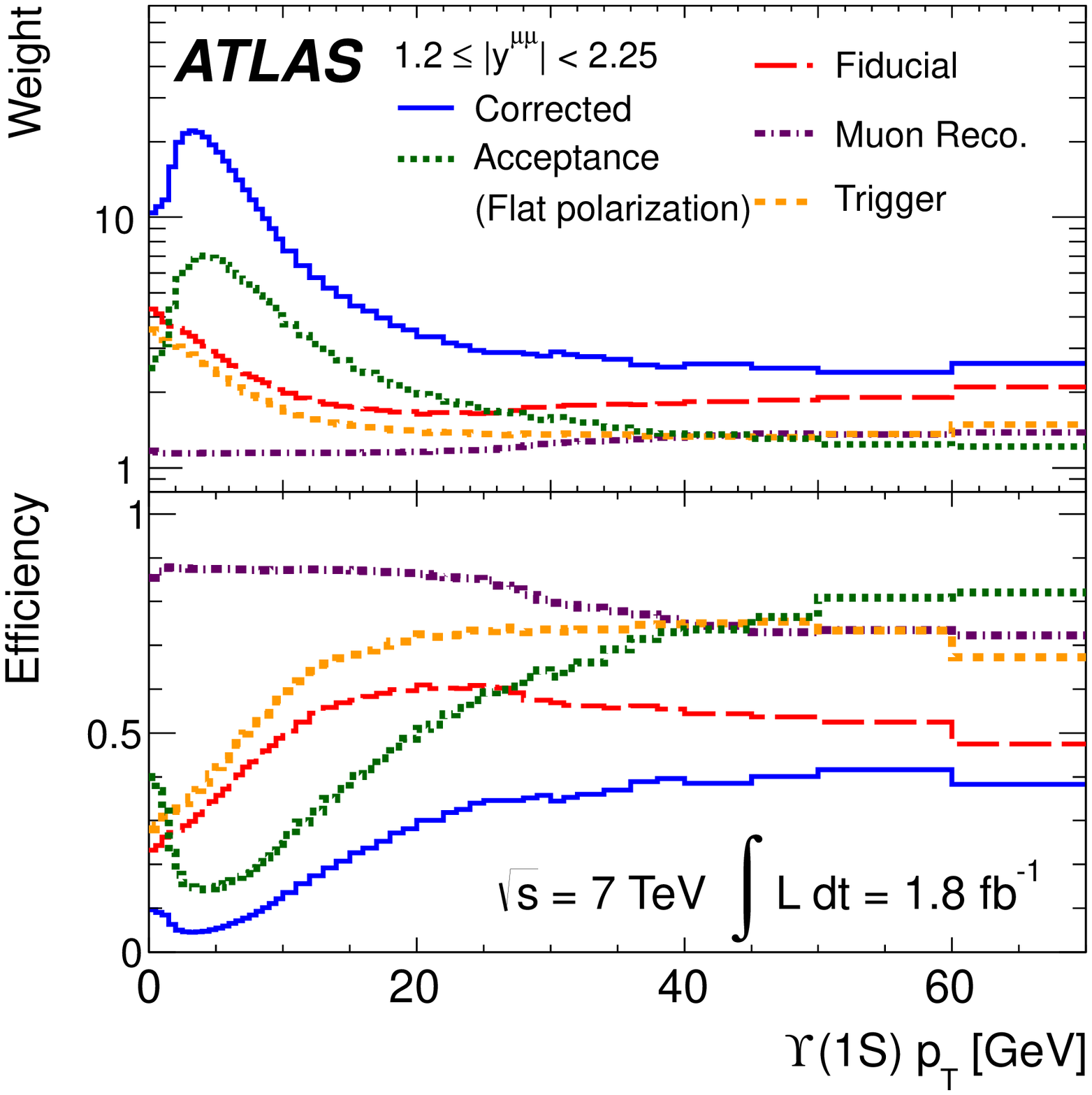}
    \includegraphics[width=0.42\textwidth]{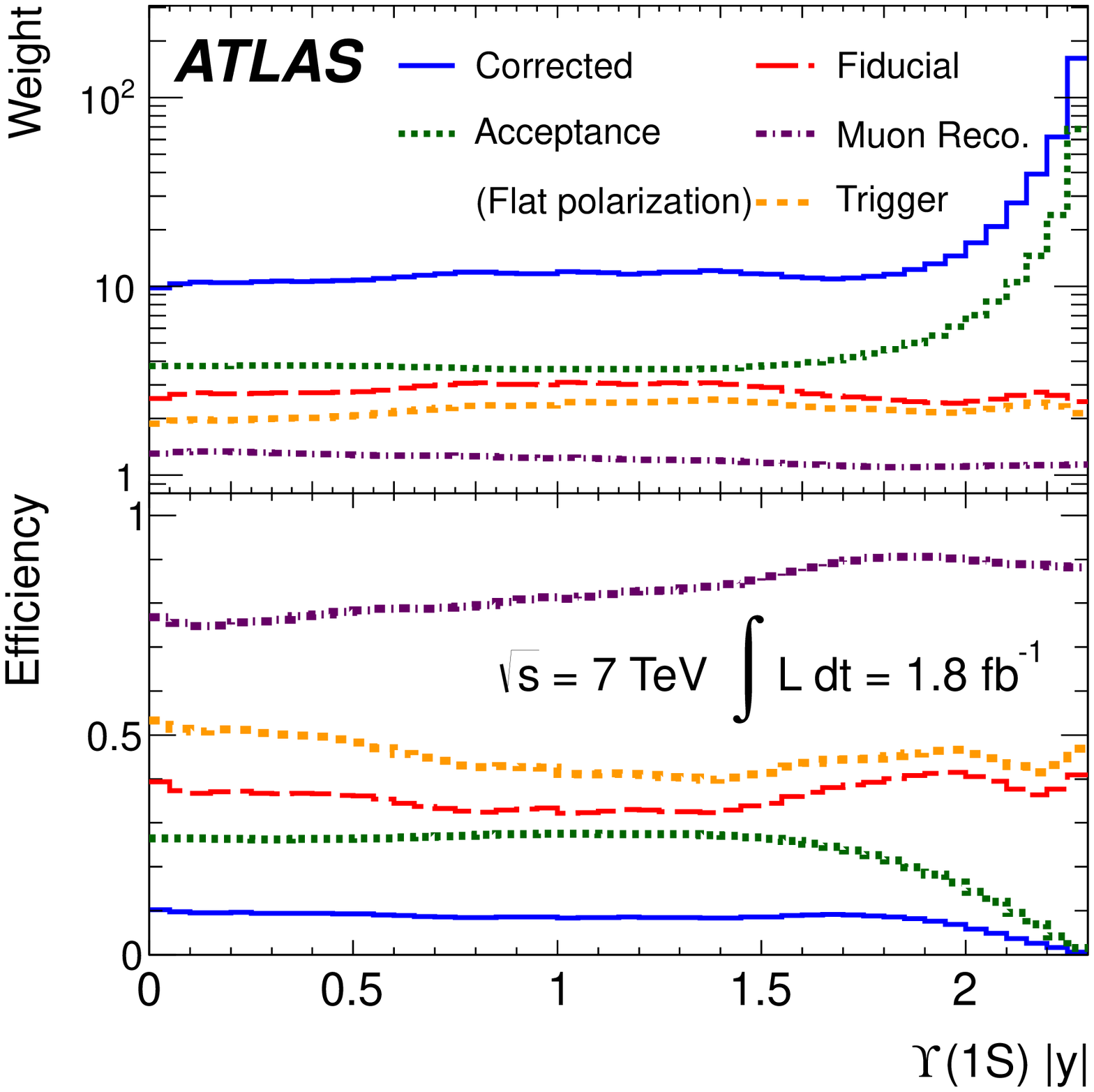}
  \caption{Efficiency and acceptance corrections contributing to the overall fiducial and corrected efficiencies
    (and the inverse of the efficiency, the weight) as a function of the $\varUpsilon(1\mathrm{S})$ \pTmumu\ in the central
    (top) and forward (middle) regions of the detector and as a function
    of \absymumu\ (bottom).}
  \label{fig:eff_main}
\end{center}
\end{figure}

The individual and overall efficiencies and weight corrections, calculated using the methods described
above, are shown as functions of \pTmumu\ and \absymumu\ in Fig.~\ref{fig:eff_main} for the $\varUpsilon(1\mathrm{S})$.

\subsection{Extracting the number of $\varUpsilon$ mesons}
\label{sect:fits}
The number of produced \UpsnS\ mesons used in our cross-section determination is found by fitting signal and background functions to
the \Mmumu\ spectrum of weighted \Ups\ candidates. The use of per-candidate weights rather than average weights allows us to correct for 
acceptance and efficiency without introduction of biases associated with the use of average values for these quantities. 
We determine the $\varUpsilon$ differential cross sections
separately for each spin-alignment scenario.
We perform least squares fits to \Mmumu\ histograms filled using the weights, 
$w$, for each candidate in bins of di-muon transverse momentum, \pTmumu, and rapidity, \ymumu.
The form of the \chisq\ for each (\pTmumu, \absymumu) bin is:
\begin{align}
\begin{split}
  \chisq &=  \sum_{i = \Mmumu^\mathrm{bin}} \left(  \frac{n_i^{\mathrm{data}} - n_i^{\mathrm{pred}}}{\sigma_i}  \right)^2 \\
  n_i^{\mathrm{data}} &=  \sum_{j = N_i} w_j; \qquad
  \sigma_i^2 =  \sum_{j = N_i} w_j^2 
\end{split}
\end{align}
where $N_i$ represents the number of \UpsnS\ candidates in bin $i$.

Predictions for each bin are constructed using four sources of
di-muon candidates:
\begin{align}
\begin{split}
  n^{\mathrm{pred}}(\Mmumu ) 
 = & \NXS{1} F_{1S}(\Mmumu ; f_{1S})\\
   & + \NXS{2} F_{2S}(\Mmumu ; f_{2S})\\
   & + \NXS{3} F_{3S}(\Mmumu ; f_{3S})\\
   & + F_{\mathrm{bgd}}(\Mmumu ; f_{\mathrm{bgd}})
\end{split}
\end{align}
where the ${f}_X$ here represent fit function parameters
and the $F_X$ are normalized probability density functions.
In order to avoid bias due to the choice of fit model, several
parameterizations of signal and background, which describe the data well,
are used. 

\begin{figure*}[htbp]
\begin{center}
    \includegraphics[width=0.43\textwidth]{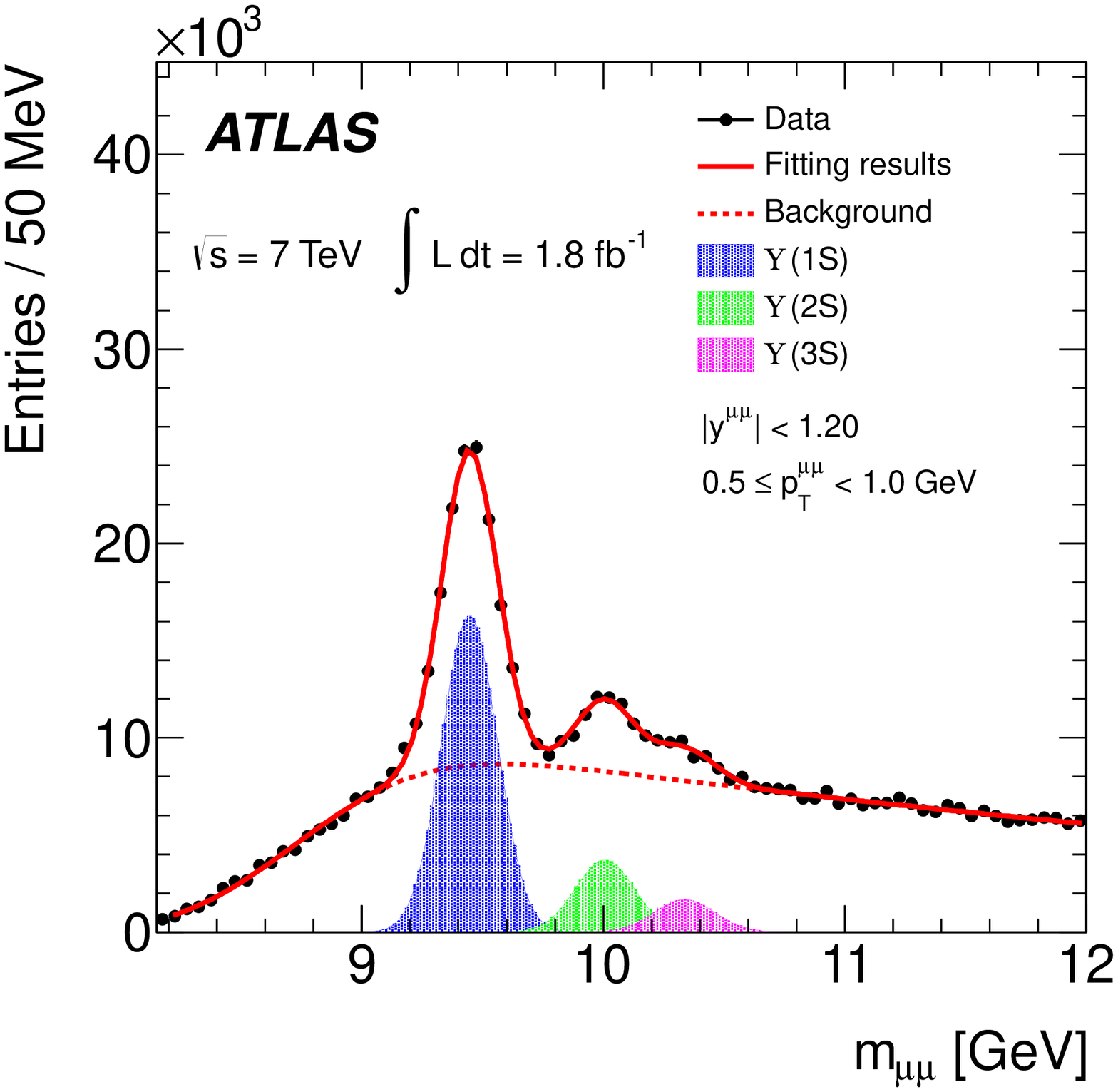}
    \includegraphics[width=0.43\textwidth]{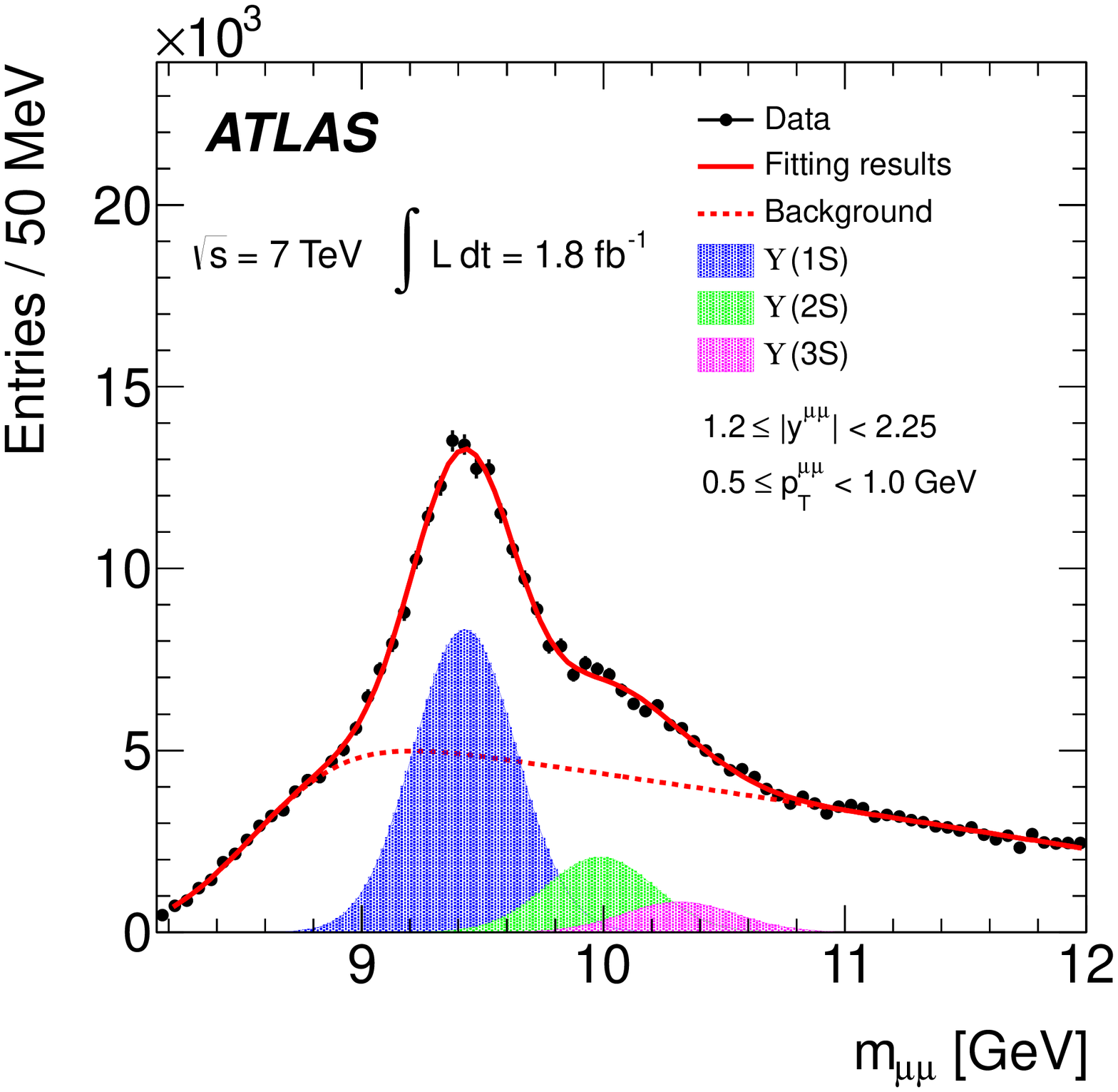}
    \includegraphics[width=0.43\textwidth]{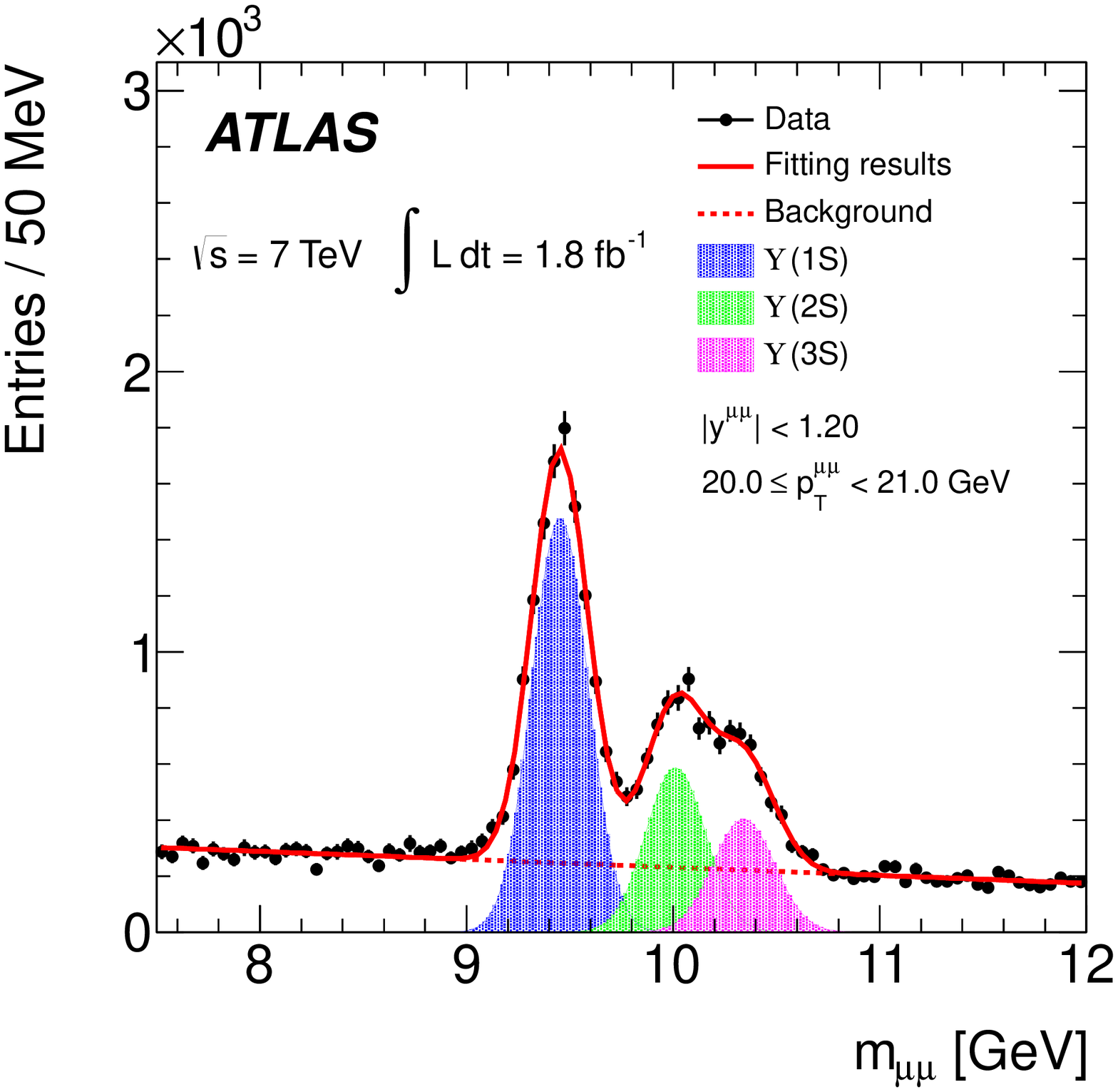}
    \includegraphics[width=0.43\textwidth]{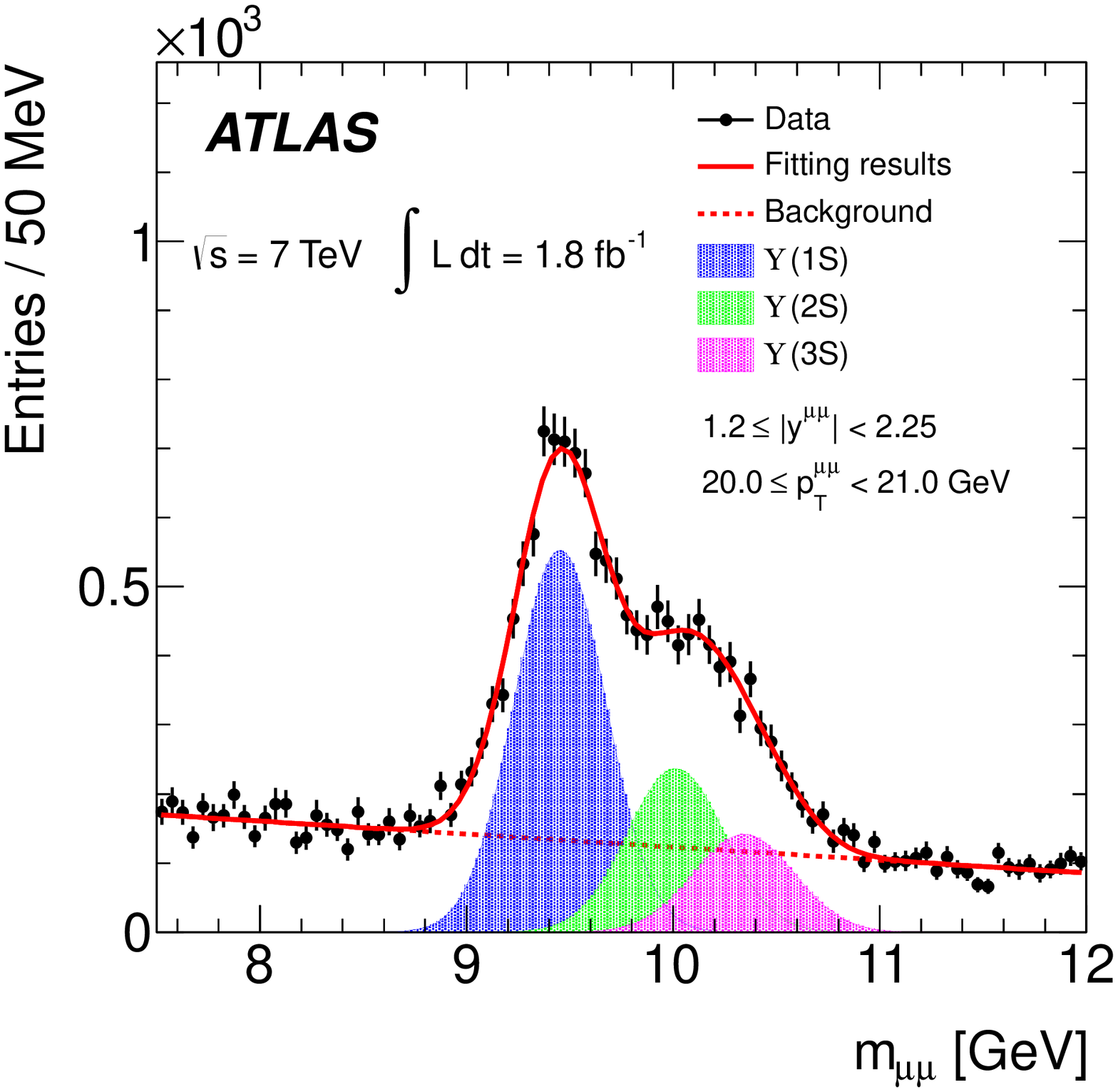}
    \includegraphics[width=0.43\textwidth]{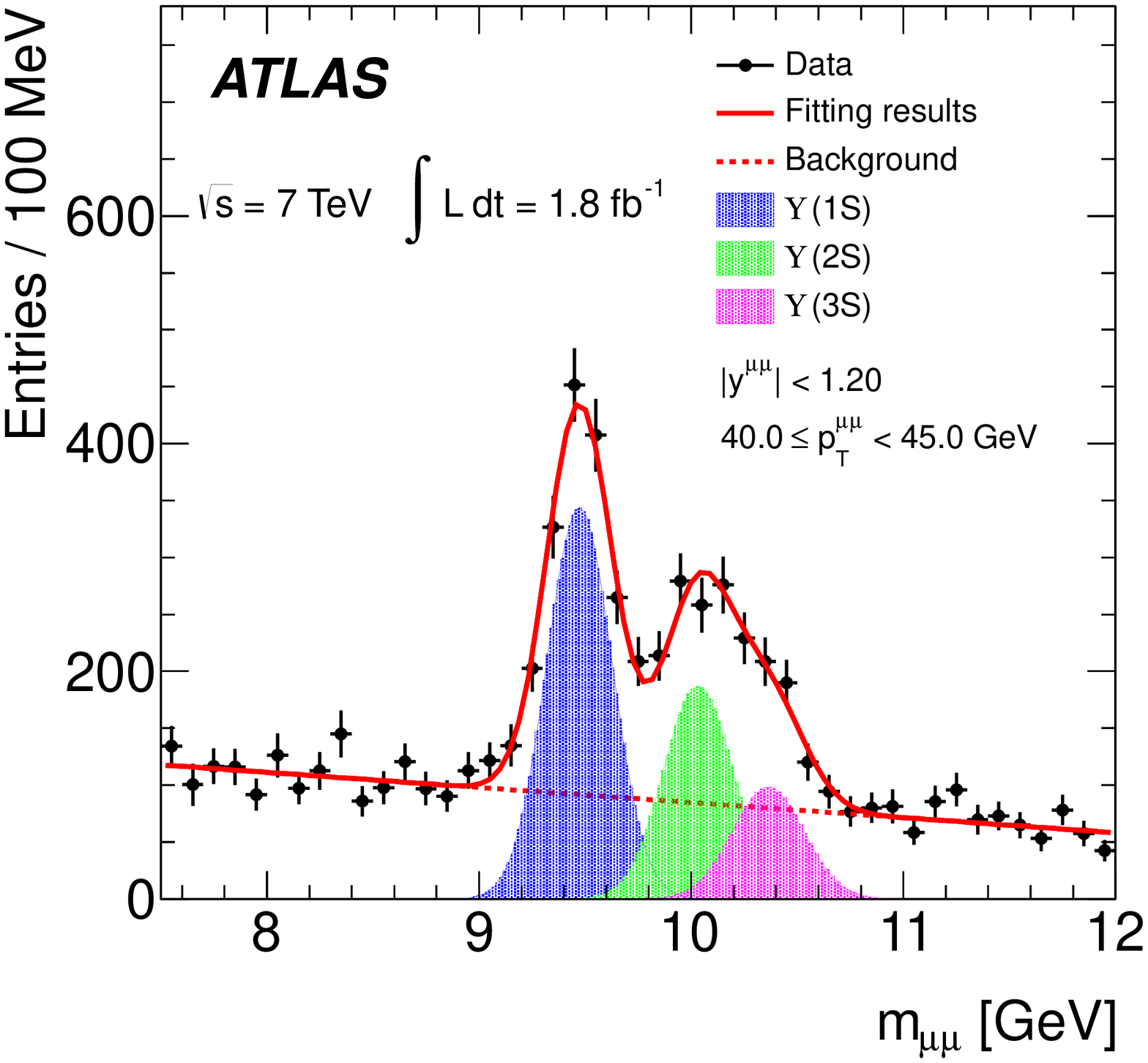}
    \includegraphics[width=0.43\textwidth]{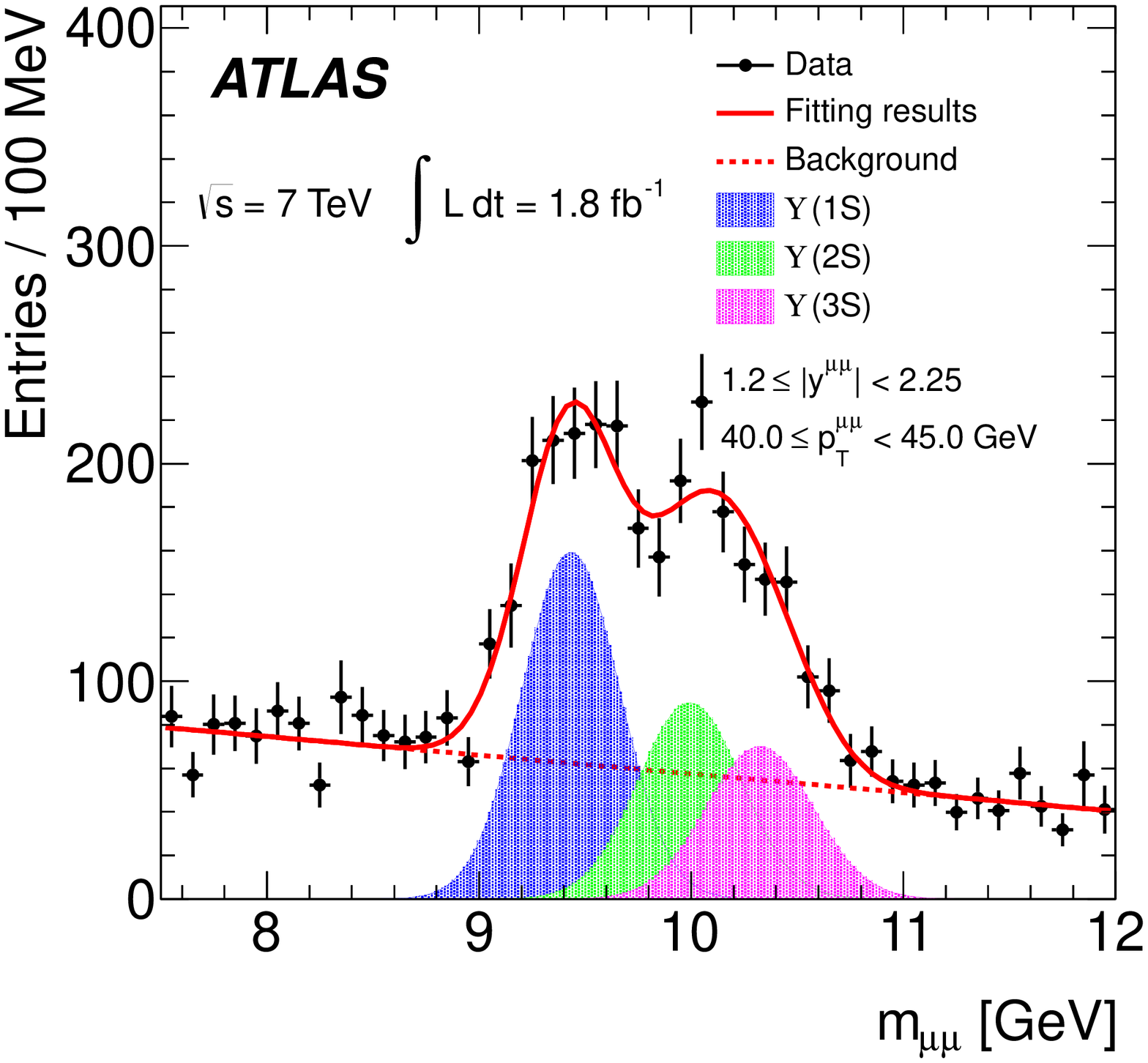}
  \caption{Fits to the efficiency-corrected \Mmumu\ spectra for candidates in the low (0.5--1.0\,\GeV; top), mid (20--21\,\GeV; middle), and high (40--45\,\GeV; bottom) \pTmumu\ intervals,
  for central (left) and forward (right) di-muon rapidities.
  Fit results are shown for the simplest fit model considered: a single Gaussian signal plus either a second-order polynomial multiplied by an error function (low $p_{\rm T}$),
  second-order polynomial (mid $p_{\rm T}$), or first-order polynomial (high $p_{\rm T}$) background parameterization.
  }
  \label{fig:massdistribution}
\end{center}
\end{figure*}

Each of the three $\varUpsilon$ resonances are parameterized by single Gaussian, 
double Gaussian, or Crystal Ball functions, chosen to provide a reasonable description of the
experimental mass resolution and energy loss effects that dominate the observed signal line-shapes.
The background parameterizations vary with di-muon \pT. 
At low \pTmumu, and for all rapidity bins, an error function multiplied by either a second-order polynomial, 
or a second-order polynomial plus an exponential, is used to model \Mmumu\ turn-on effects accurately. 
At mid \pTmumu\ a second-order polynomial, or second-order polynomial plus an exponential
is adequate to describe the shape of the background under the $\varUpsilon$ peaks.
At high \pTmumu\ a first-order polynomial or a first-order polynomial plus an exponential is sufficient to describe the background contribution.
Average fitted values for the numbers of \UpsnS\ mesons (\NXS{1}, \NXS{2}, \NXS{3}) taken over all combinations of signal and
background models are used in the extraction of cross sections, while
the maximum deviation of any fit result from the average is used as
an estimate of the fit model systematic uncertainty.

In the fits \NXS{1}, \NXS{2}, \NXS{3}, the peak mass of the \UpsXS{1} meson (\MXS{1}),
parameters describing the shape of the \UpsMuMu\ mass distribution, and the background parameters are allowed to vary freely while 
\MXS{2}, \MXS{3} are fixed relative to \MXS{1} using measured mass splittings\,\cite{PDG};
and the line-shape width parameters \SigXS{2}, \SigXS{3} are related linearly in \Mmumu\ to \SigXS{1}. 
The distribution of the \chisq-probabilities from the fits are uniformly distributed indicating that our signal and background parameterizations 
provide good descriptions of the data.
Figure~\ref{fig:massdistribution} shows the invariant mass distribution of the fitted signal region
for representative low, mid, and high \Ups\ $p_{\rm T}$ intervals, in both the central and forward rapidity regions.

In order to account for bin migrations due to finite detector resolution,
corrections in \pTmumu\ are derived by first fitting a smooth analytic function
to the background-subtracted \pTmumu\ spectra of $\varUpsilon$ events in data. This fitted distribution is 
deconvolved with a Gaussian distribution using a Gauss-Hermite quadrature integration technique,
with a $p_{\rm T}$ resolution derived from the fitted invariant mass and muon angular resolutions in data.
From the distributions with and without convolution, migration corrections bin-by-bin are derived. 
These corrections are found to be within 0.5\% of unity for 
central rapidities and within 1\% of unity for forward rapidities. 
The number of efficiency-corrected \UpsMuMu\ decays extracted from
the fits in each (\pTmumu, \ymumu) bin are corrected for the difference
between true and reconstructed values of the di-muon \pT\ and rapidity.
Using a similar technique, bin-migration corrections as functions of \ymumu\ are found 
to be negligible.

\section{Systematic Uncertainties}
\label{sect:Systematics}
\begin{figure*}[htbp]
\begin{center}
  \includegraphics[width=0.85\textwidth]{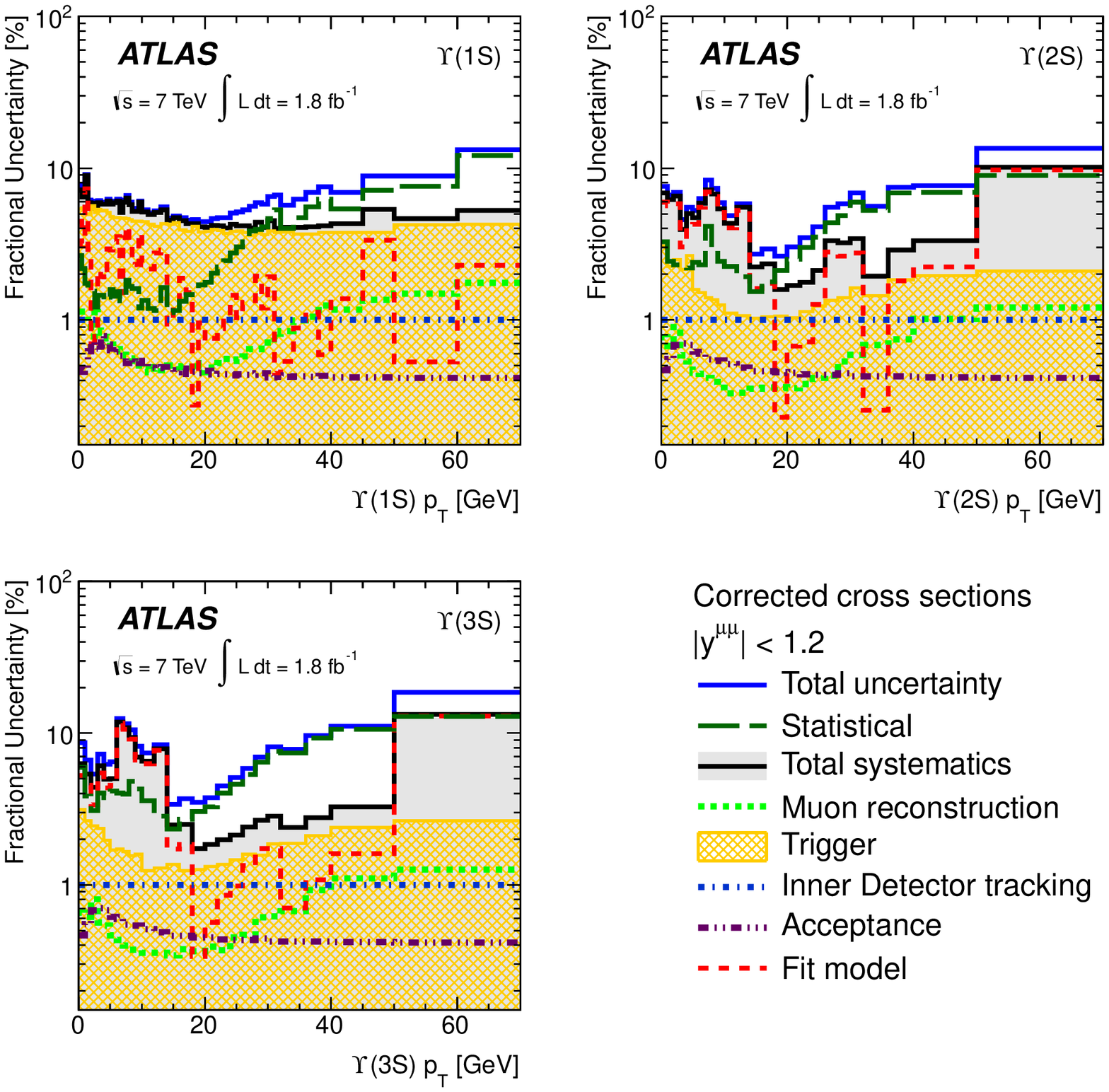}
  \caption{Sources of statistical and systematic uncertainty on the corrected $\varUpsilon(n\mathrm{S})$ production
    cross-section measurements in the central rapidity region. A common luminosity uncertainty of $\pm3.9\%$ is not included.}
  \label{fig:SystTot-c}
\end{center}
\end{figure*}

\begin{figure*}[htbp]
\begin{center}
  \includegraphics[width=0.85\textwidth]{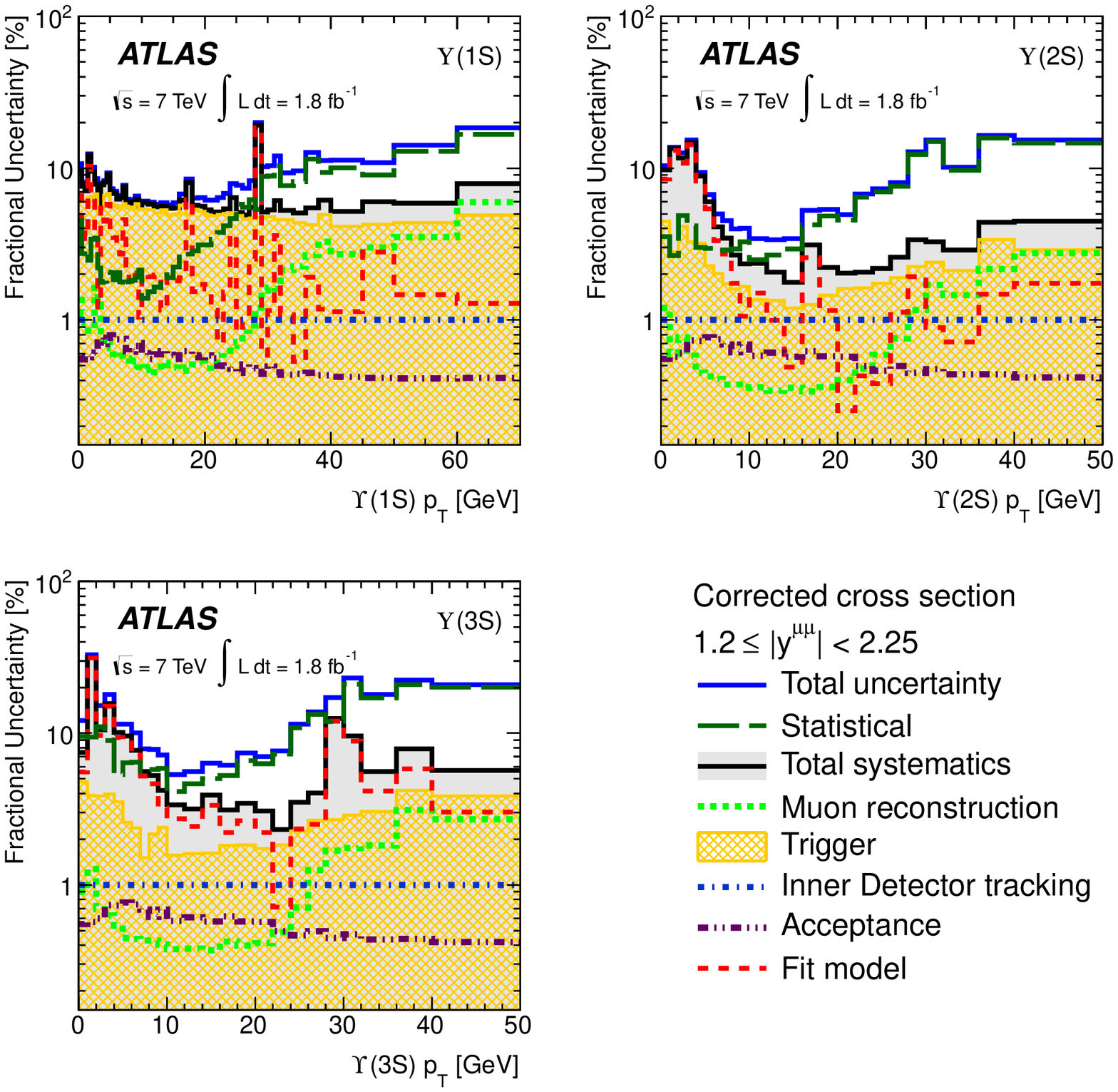}
  \caption{Sources of statistical and systematic uncertainty on the corrected $\varUpsilon(n\mathrm{S})$ production
    cross-section measurements in the forward rapidity region. A common luminosity uncertainty of $\pm3.9\%$ is not included.}
  \label{fig:SystTot-f}
\end{center}
\end{figure*}

\begin{figure*}[htbp]
\begin{center}
  \includegraphics[width=0.85\textwidth]{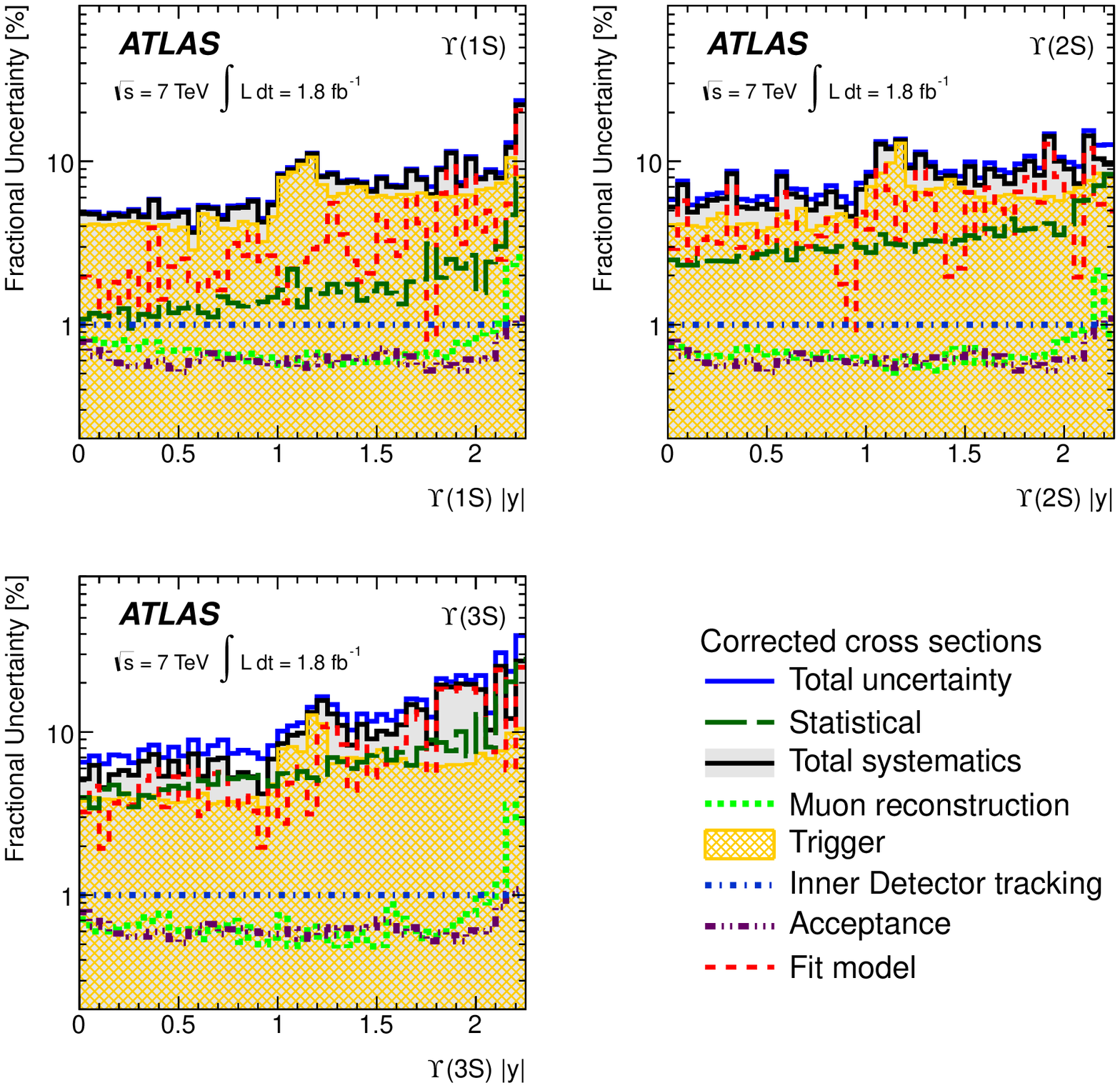}
  \caption{Sources of statistical and systematic uncertainty on the corrected $\varUpsilon(n\mathrm{S})$ production
    cross-section measurements as a function of \Ups\ rapidity. A common luminosity uncertainty of $\pm3.9\%$ is not included.}
  \label{fig:SystTot-y}
\end{center}
\end{figure*}

We consider the following sources of systematic uncertainty on
the \UpsnS\ differential cross sections:
luminosity determination, reconstruction and trigger efficiencies, migration between
\pTmumu\ and \absymumu\ bins due to resolution, acceptance corrections, and the background and signal fit models used. 
The range of these uncertainties for the three $\varUpsilon$ states is summarized in
Table~\ref{table:syst} and their breakdown in each source is given for the
corrected cross-section analysis in Figs.~\ref{fig:SystTot-c}--\ref{fig:SystTot-y}. 

\setlength{\tabcolsep}{12pt}
\renewcommand{\arraystretch}{1.1}
\begin{table*}[htbp]
\caption{Summary of statistical and systematic uncertainties on the cross-section
  measurements. 
  The systematic uncertainty due to the acceptance determination applies only to the corrected
  cross-section measurements and does not include possible variation of the result
  due to spin-alignment.
  Values quoted refer to the range of uncertainties over \pTmumu\ in
  each of the \ymumu\ regions considered.
}
\label{table:syst}
\begin{center}
\begin{tabular}{lcc}
\hline \hline
  & \multicolumn{2}{c}{Relative Uncertainty
    [\UpsXS{1} / \UpsXS{2} / \UpsXS{3}] [\%]}\\
  Source & $\absymumu < 1.2$ & $1.2\leq \absymumu < 2.25$\\
\hline
  Statistical Uncertainty & 1.5--10 / 2--10 / 3--10 & 2--15 / 2.5--12 / 4--18 \\
  Fit Model & 0.3--2 / 0.3--10 / 0.3--15   & 0.4--10 / 0.3--5.0 / 0.2--12\\
  Luminosity & 3.9 & 3.9\\
  Trigger Efficiency & 2.0--3.0 / 1.0--2.0 / 1.0--2.0   & 2.0--3.0 / 2.0--4.0 / 2.0--7.0\\
  Muon Reconstruction Efficiency & 0.7--2.0 / 0.4--1.2 / 0.4--1.2    & 0.5--5.0 / 0.4--3.0 / 0.4--3.0\\
  Acceptance corrections & 0.7--1.5 & 0.7--1.5\\
  Track Reconstruction Efficiency & 1.0 & 1.0 \\
\hline \hline
\end{tabular}
\end{center}
\end{table*}

The relative luminosity uncertainty of 3.9\%
is described in more detail in Ref.\,\cite{ATLAS:lumi}. Other
sources of systematic uncertainty are discussed below.
As the statistical components of the uncertainties associated with the determination of \effReco\ and \effTrig\ are dominant, the uncertainties on the cross sections 
are derived from the statistics of the control samples used to extract them using a series of pseudo-experiments,
randomly varying the weights used for each \Ups\ candidate in data, according to the
uncertainties on the efficiency maps.
Systematic uncertainties associated with the fit model used to extract
the number of \UpsnS\ decays from our di-muon data sample, are
quantified by taking the largest deviation of the fitted values of \NXS{n}
found in the six possible model combinations from the average value. 
Systematic effects, due to differences in the underlying kinematic
distributions of the control sample and the data distributions,
are found to be negligible due to the fine differential binning of the efficiency and acceptance corrections.

Uncertainties associated with the acceptance correction include statistical uncertainties on
the determination of the correction in fine $p_{\rm T}$ and rapidity bins. This constitutes approximately a 0.5\%
uncertainty across the measured spectrum. 
A shift in the interaction point along the beamline axis, $z$, influences the acceptance particularly at large rapidity. 
To estimate the impact of this effect, the acceptances were recalculated with shifts in $z$ by $\pm 62$\,mm, corresponding
to one Gaussian standard deviation of the vertex $z$ position distribution in the analyzed data, and the variation in the measured cross sections was calculated
as a function of $\varUpsilon$ $p_{\rm T}$ and rapidity. These variations result in 
uncertainties on the acceptance corrections of $0.4$--$0.7\%$ as a function of $p_{\rm T}$, and $0.6$--$1.5\%$ as a function
of rapidity, growing toward larger rapidities and lower transverse momenta.
Our estimate of the correction to the number of \UpsMuMu\ decays fitted in each (\pTmumu, \ymumu) bin for
differences between reconstructed and true values of \pTmumu\ due to bin migration depends on a knowledge of the di-muon
\pT\ resolution. Allowing this resolution to vary within its uncertainty results in a negligible change in the
\UpsnS\ cross sections. 

As mentioned earlier, \Ups\ spin-alignment
effects have an impact on the determination of cross sections
extrapolated to full phase space. As in the ATLAS measurement
of \Jpsi\ production\,\cite{ATLAS:jpsi}, we quote an uncertainty due to
\Ups\ spin-alignment by comparing results obtained using the
unpolarized assumption 
($\lambda_{\theta} = \lambda_{\phi} = \lambda_{\theta \phi} = 0$) to
those derived under the other four extreme parameter combinations. The size of the
possible variation in each bin is taken to be the largest positive and
negative deviation from the unpolarized baseline result.
We provide the cross-section results for each of the spin-alignment scenarios
along with tabulated values\,\cite{HEPDATA} of the weight corrections 
so that the corrected data can be determined in terms of any spin-alignment scenario.

\section{Results and Discussion}
\label{sect:Results}
We measure the differential cross section multiplied by the di-muon branching fractions
of $\varUpsilon(1\mathrm{S})$, $\varUpsilon(2\mathrm{S})$, and $\varUpsilon(3\mathrm{S})$ mesons as a function of $\varUpsilon$ transverse momentum and rapidity,
both in a fiducial region defined by $p^{\mu}_{\rm T}>4$\,\GeV, $|\eta^{\mu}|<2.3$ (free from spin-alignment uncertainties), 
and corrected back to the full muon decay phase space with $|y^\varUpsilon|<2.25$.
We additionally present measurements of the production cross sections of the \UpsXS{2} and \UpsXS{3} 
relative to the \UpsXS{1} as a function of \UpsXS{1} transverse momentum and rapidity.
Tabulated results of all measurements presented in this paper are available in \textsc{hepdata}\,\cite{HEPDATA}.

\subsection{Fiducial cross sections}
\label{sec:fiducialresults}
\begin{figure}[htbp]
\begin{center}
    \includegraphics[width=0.46\textwidth]{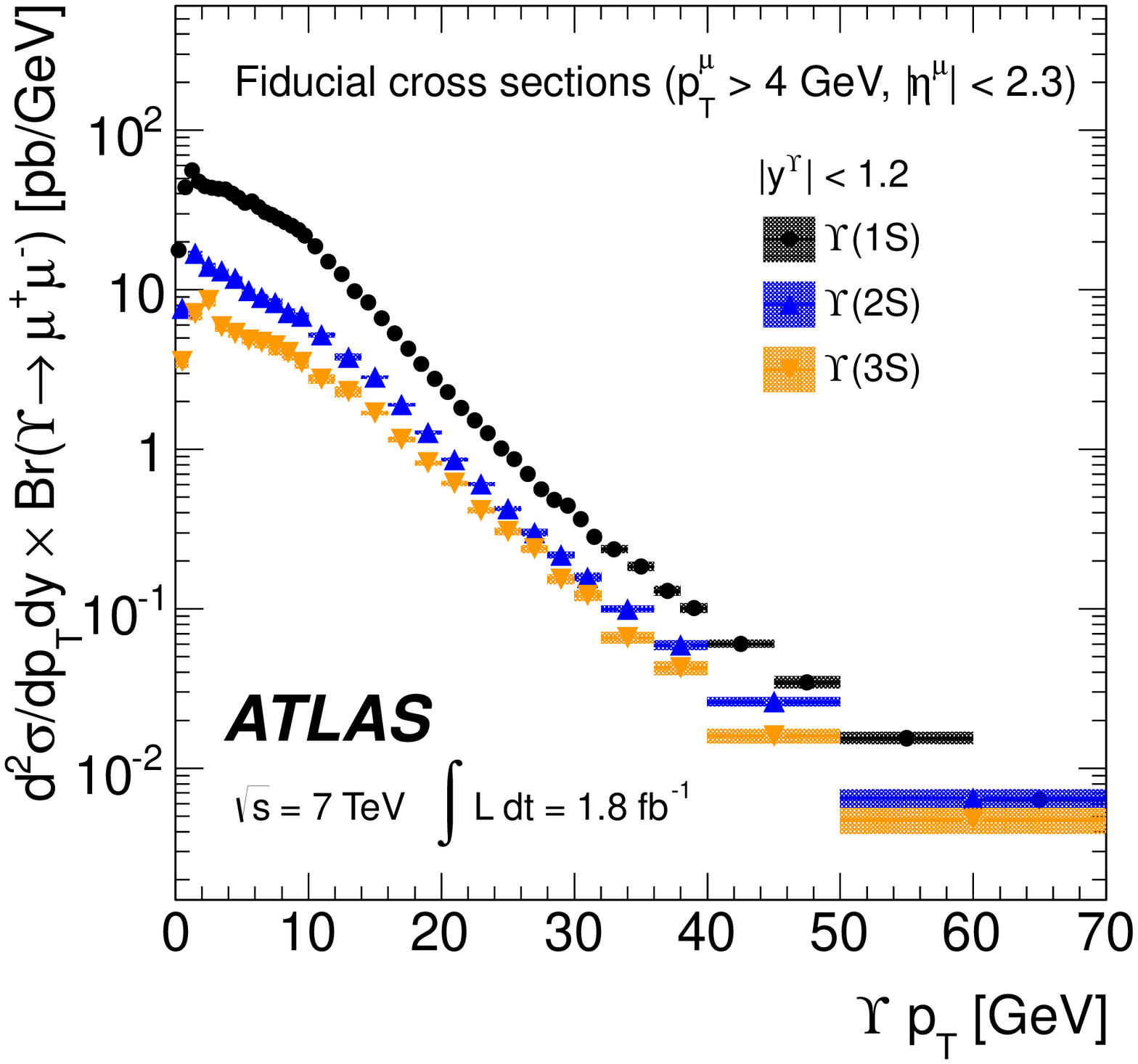}
    \includegraphics[width=0.46\textwidth]{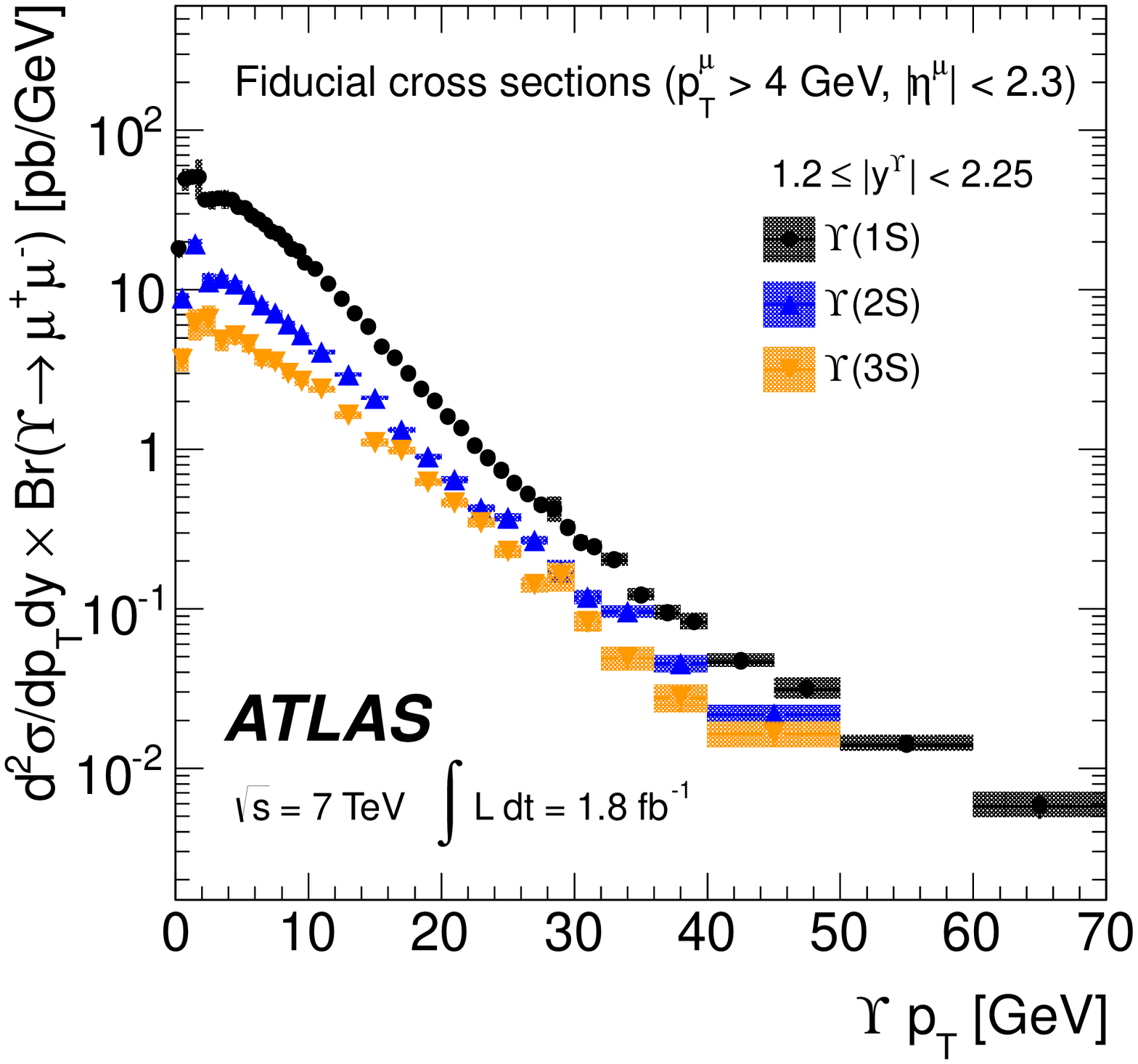}
  \caption{Differential cross sections multiplied by the  di-muon branching fraction,
    \XSpTyB , for central (top) and forward (bottom) rapidities,
    for \UpsXS{1}, \UpsXS{2}, \UpsXS{3} production within the fiducial
    acceptance.
    Points with error bars indicate results of the measurements with
    statistical uncertainties while shaded areas correspond to total uncertainties
    on the measurement, including systematic effects.
  }
  \label{fig:FidXS}
\end{center}
\end{figure}

\begin{figure}[htbp]
\begin{center}
    \includegraphics[width=0.46\textwidth]{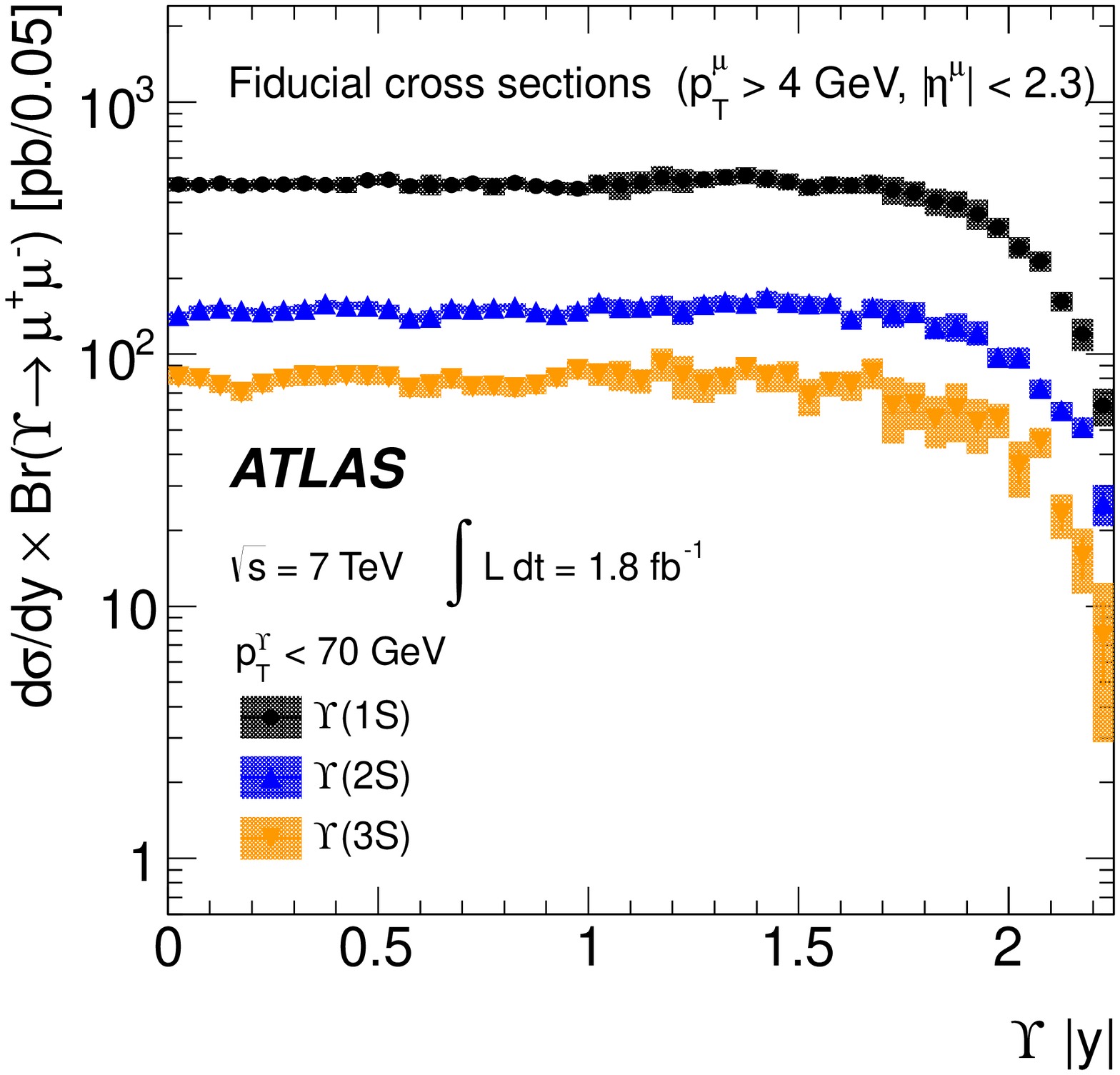}
  \caption{Differential cross sections multiplied by the  di-muon branching fraction,
    \XSyB ($p_{\rm T}$-integrated) \XSyB
    for \UpsXS{1}, \UpsXS{2}, \UpsXS{3} production within the fiducial
    acceptance.
    Points with error bars indicate results of the measurements with
    statistical uncertainties while shaded areas correspond to total uncertainties
    on the measurement, including systematic effects.
  }
  \label{fig:FidXS-y}
\end{center}
\end{figure}

Differential cross sections multiplied by the \UpsMuMu\ branching fractions,
\XSpTyB, are calculated within the fiducial acceptance of our analysis
($p^{\mu}_{\rm T}>4$\,\GeV, $|\eta^{\mu}|<2.3$)
using the results of fits to di-muon candidates in data 
corrected, candidate by candidate, for efficiencies using the fiducial
weights, \wFid, defined in Eq.~(\ref{eqn:weight}).
These differential fiducial cross sections along with total uncertainties are shown in Figs.~\ref{fig:FidXS} and~\ref{fig:FidXS-y}
and span four orders of magnitude across the $p_{\rm T}$ range studied. 

Integrating over all \pTUps\ and both rapidity bins we find cross sections within the fiducial acceptance of the detector as shown
in Table~\ref{table:FidXS}.
The results for the \UpsXS{1} are consistent with our previous measurement\,\cite{ATLAS:upsilon} but extend to a significantly higher \pTUps\ 
and with increased precision. 

\begin{table}[htbp]
\caption{Integrated fiducial cross-section measurements for $\varUpsilon(n\mathrm{S})$. Uncertainties quoted
  represent statistical, systematic, and luminosity terms respectively.
}
\label{table:FidXS}
\begin{center}
\begin{tabular}{cc}
\hline \hline
\multicolumn{2}{c}{Integrated fiducial cross sections}\\
\multicolumn{2}{c}{ $\pTmu > 4$\,\GeV, $\etamu < 2.3$} \\
\hline
  State 
    & $\sigma_\mathrm{fid}(pp \to \Ups)\times \mathrm{Br}(\Ups \to \mu^+ \mu^-)$\\
  & Range: $\pTUps<70$\,\GeV, $|y^\varUpsilon|<2.25$\\
\hline
  \UpsXS{1} & 1.890 $\pm$ 0.007 $\pm$ 0.095 $\pm$ 0.074 nb\\
  \UpsXS{2} & 0.601 $\pm$ 0.003 $\pm$ 0.040 $\pm$ 0.023 nb\\
  \UpsXS{3} & 0.304 $\pm$ 0.003 $\pm$ 0.021 $\pm$ 0.012 nb\\
\hline \hline
\end{tabular}
\end{center}
\end{table}
Presenting the results in a restricted kinematic phase space removes any uncertainty due to the spin-alignment
of quarkonia. This allows unambiguous comparison of differential spectra with any theoretical approaches which can provide 
predictions with kinematic restrictions applied to the decay products of the $\varUpsilon$. 

NLO color-singlet\,\cite{CSM} predictions\,\cite{NLOsinglet} have previously been compared to
differential $\varUpsilon(1\mathrm{S})$ fiducial production cross sections\,\cite{ATLAS:upsilon} and were found 
to underestimate the measured production rates by approximately an order of magnitude
and to not reflect the $p_{\rm T}$ dependence of the data.
This is not surprising as it has been known\,\cite{NNLOUpsilon2} for some time that higher-order corrections are both large
and necessary in order to adequately describe quarkonium production at high $p_{\rm T}$ with color-singlet terms. 
At this time, no color-singlet calculations beyond NLO are available for quarkonium production measurements quoted in a restricted muon acceptance,
nor are Color Octet\,\cite{COM} or Color Evaporation\,\cite{CEM} approaches currently able to account for the kinematics of leptons from the quarkonium decay.
The fiducial measurements presented here are precise and free from any assumptions on the angular dependencies of the di-muon system, accurately
reflecting the production dynamics in proton-proton collisons at $\sqrt{s}=7$\,TeV.

\subsection{Corrected cross sections}
\label{sec:inclusiveresults}
\begin{figure}[phtb]
  \begin{center}        
    \includegraphics[width=0.46\textwidth]{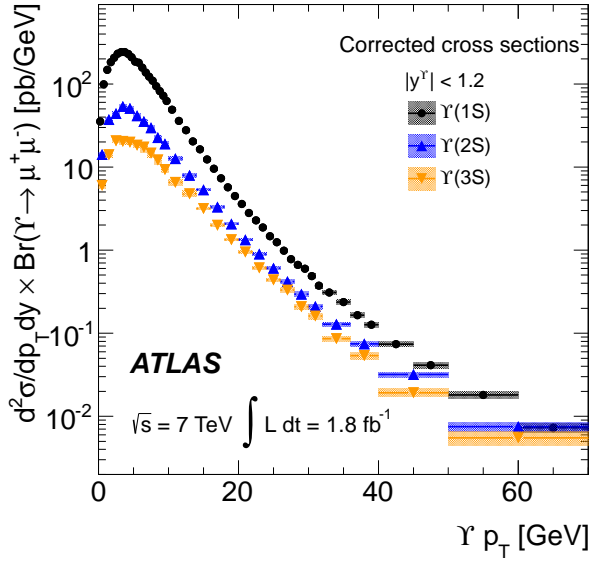}
    \includegraphics[width=0.46\textwidth]{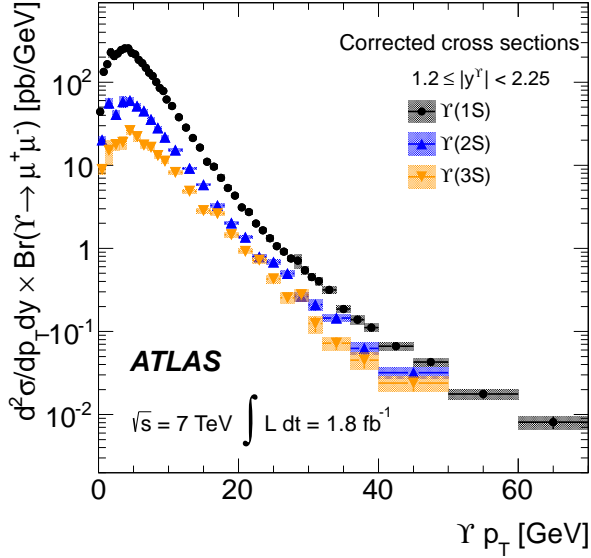}
  \caption{Differential cross sections multiplied by the  di-muon branching fraction,
    for \UpsXS{1}, \UpsXS{2} and \UpsXS{3} production extrapolated to the
    full phase space for the (top) $|y^\varUpsilon|<1.2$, (bottom) $1.2\leq|y^\varUpsilon|<2.25$ rapidity intervals.
    Points with error bars indicate the results of the measurements with
    total statistical and systematic errors. Results are shown assuming an isotropic spin-alignment scenario.
  }
  \label{fig:TotXS-pt_overlay}    
  \end{center}
\end{figure}

\begin{figure}[phtb]
  \begin{center}        
    \includegraphics[width=0.46\textwidth]{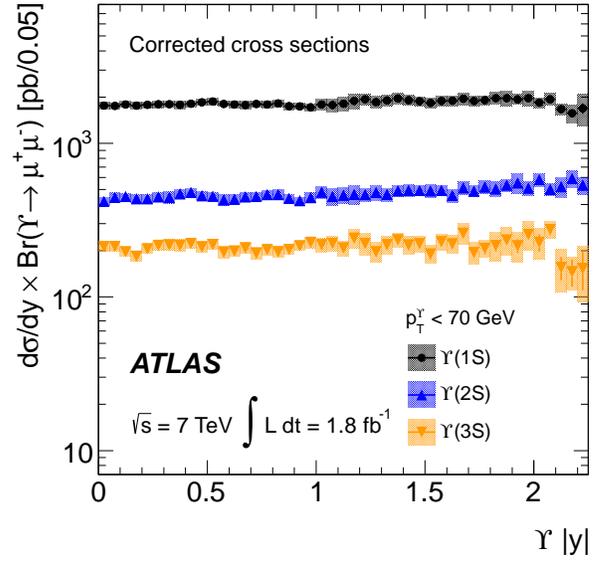}
  \caption{Differential cross sections multiplied by the  di-muon branching fraction,
    for \UpsXS{1}, \UpsXS{2} and \UpsXS{3} production extrapolated to the
    full phase space,
    $p_{\rm T}$-integrated as a function of absolute $\varUpsilon$ rapidity.
    Points with error bars indicate the results of the measurements with
    total statistical and systematic errors. Results are shown assuming an isotropic spin-alignment scenario.
  }
  \label{fig:TotXS-y}    
  \end{center}
\end{figure}

We also calculate differential cross sections multiplied by the  \UpsMuMu\ branching fractions
extrapolated to the full muon phase space within $|y^\varUpsilon|<2.25$.
For these corrected cross sections, the results of fits to di-muon candidates in data 
are corrected, candidate by candidate, for efficiencies using the total
weights, \wTot , defined in Eq.~(\ref{eqn:weight}).
Results are shown for the isotropic spin-alignment scenario in Figs.~\ref{fig:TotXS-pt_overlay} and~\ref{fig:TotXS-y} as a function of $\varUpsilon$ $p_{\rm T}$ and $y$ for all three states and, 
integrated over all \pTUps\ and both rapidity bins, in Table~\ref{table:TotXS}.

Our results are consistent with measurements by CMS\,\cite{CMS:upsilon} and, in the small region of rapidity overlap from $2.0<|y|<2.25$, with
LHCb\,\cite{LHCb:upsilon}. 
These results allow us to test phenomenological models of \Ups\ production not just at low $p_{\rm T}$ (for complementarity with
results from the Tevatron experiments) but in a newly-probed region of significantly boosted $\varUpsilon$, where higher-order contributions become particularly important.

\begin{table}[htbp]
\caption{Corrected cross-section measurements in the isotropic spin-alignment scenario. Uncertainties quoted
  represent statistical, systematic, and luminosity terms,
  respectively.
}
\label{table:TotXS}
\begin{center}
\begin{tabular}{cc}
\hline \hline
\multicolumn{2}{c}{Integrated corrected cross sections} \\
\hline
  State
    & $\sigma(pp \to \Ups)\times \mathrm{Br}(\Ups \to \mu^+ \mu^-)$\\
  & Range: $\pTUps<70$\,\GeV, $|y^\varUpsilon|<2.25$\\
\hline
  \UpsXS{1} & \totXSoneS \\
  \UpsXS{2} & \totXStwoS \\
  \UpsXS{3} & \totXSthreeS \\
\hline \hline
\end{tabular}
\end{center}
\end{table}

\begin{figure}[phtb]
  \begin{center}        
  \includegraphics[width=0.48\textwidth]{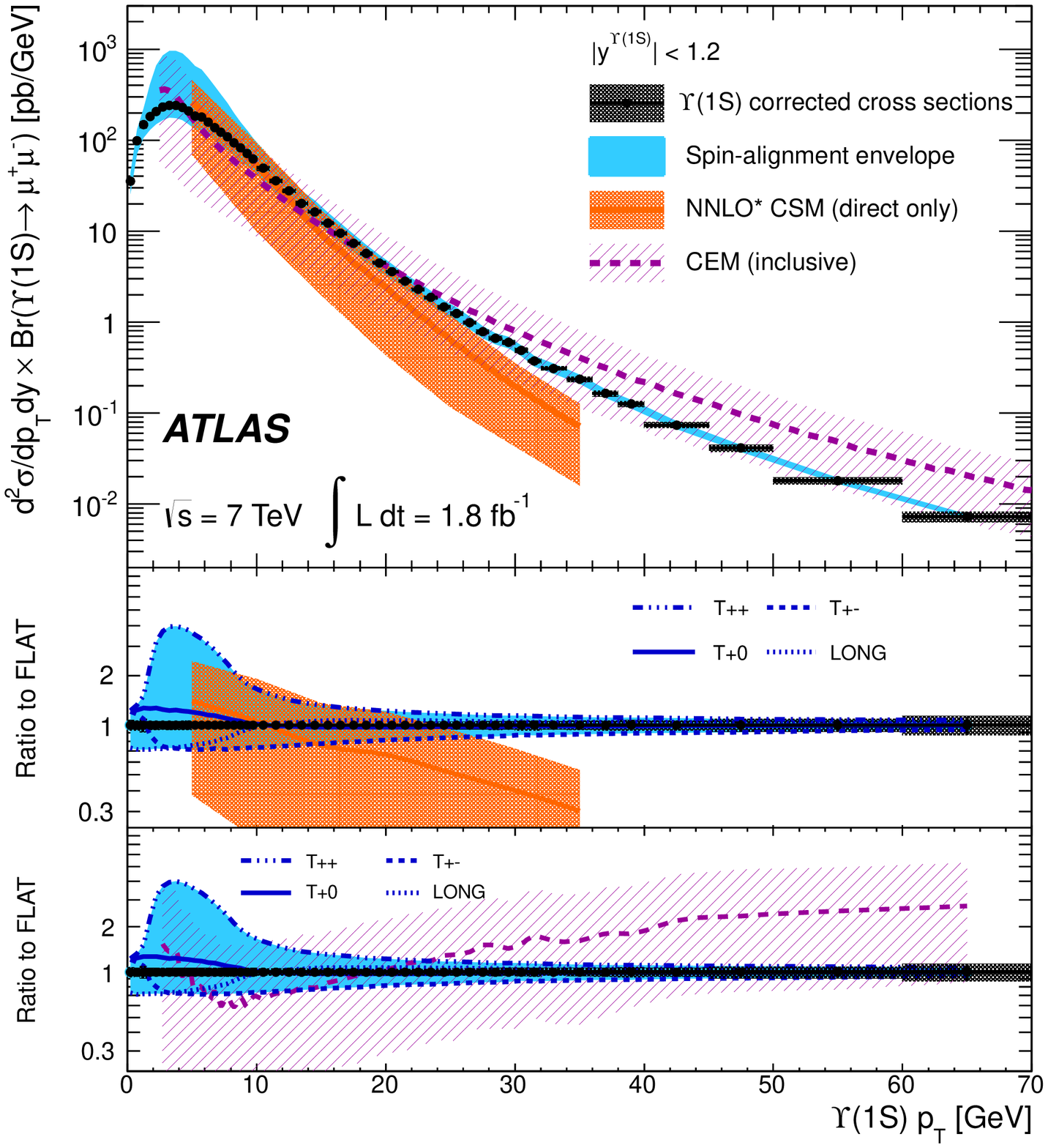}
  \includegraphics[width=0.48\textwidth]{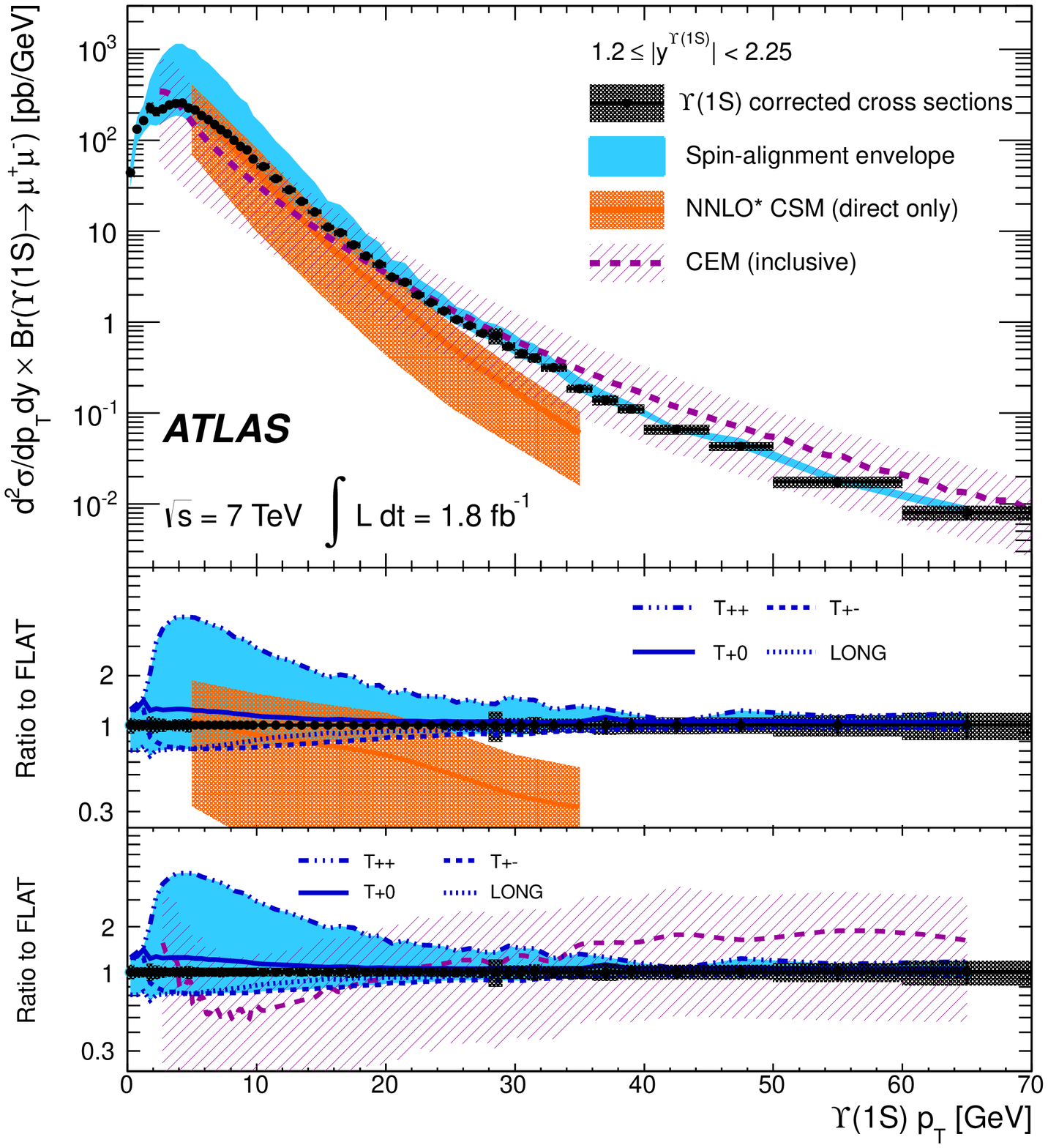}
  \caption{
    Differential cross sections multiplied by the  di-muon branching fraction,
    \XSpTyB, for \UpsXS{1} production extrapolated to the full phase space for (top) central and (bottom) forward rapidities.
    Points with error bars indicate results of the measurements with
    total (statistical and systematic) uncertainties. The maximal envelope of variation of the result due to spin-alignment uncertainty
    is indicated by the solid band.
    Also shown are 
    predictions of direct production with the NNLO* Color Singlet Mechanism (CSM) and inclusive predictions from the Color Evaporation Model (CEM).
    These theory predictions are shown as a ratio to the data in the lower panes for CEM (middle) and CSM (bottom), along with detail of the 
    variations of the cross-section measurement under the four anisotropic spin-alignment scenarios as a ratio to the nominal data.
  }
  \label{fig:TotXS-1S}    
  \end{center}
\end{figure}

\begin{figure}[phtb]
  \begin{center}        
  \includegraphics[width=0.48\textwidth]{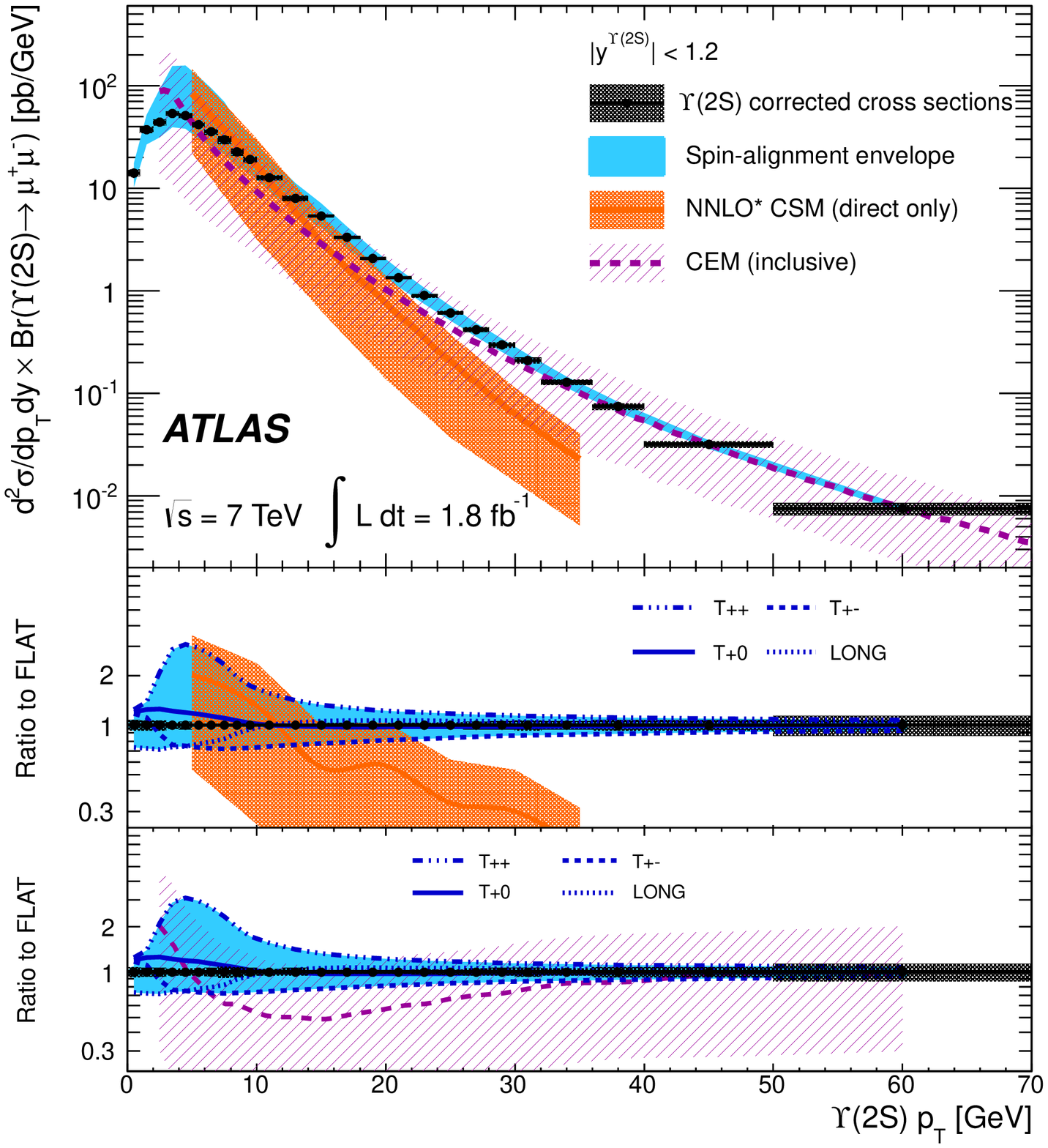}
  \includegraphics[width=0.48\textwidth]{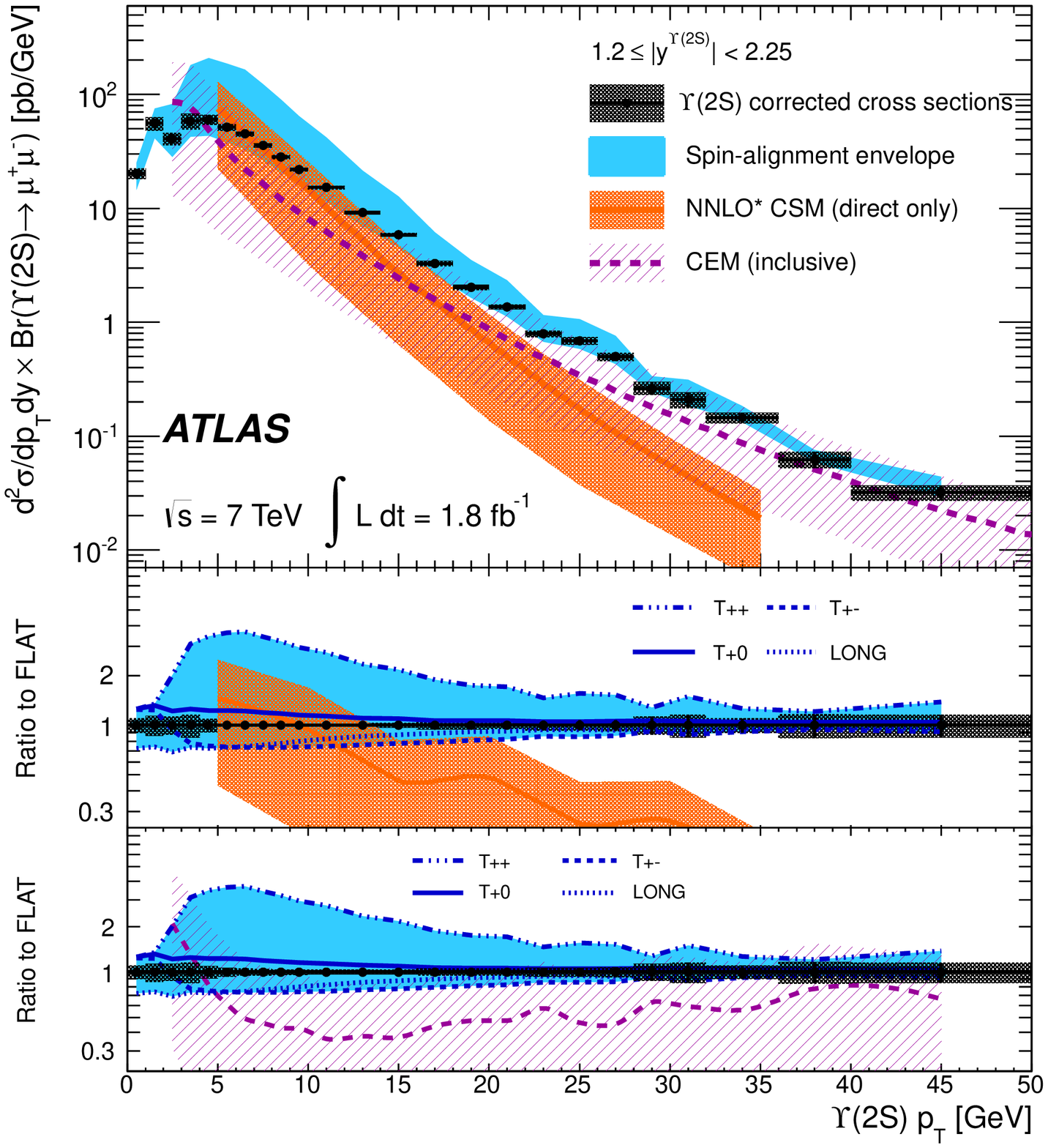}
  \caption{
  As for Fig.~\ref{fig:TotXS-1S}, but for \UpsXS{2} production. 
  }
  \label{fig:TotXS-2S}    
  \end{center}
\end{figure}

\begin{figure}[phtb]
  \begin{center}        
  \includegraphics[width=0.48\textwidth]{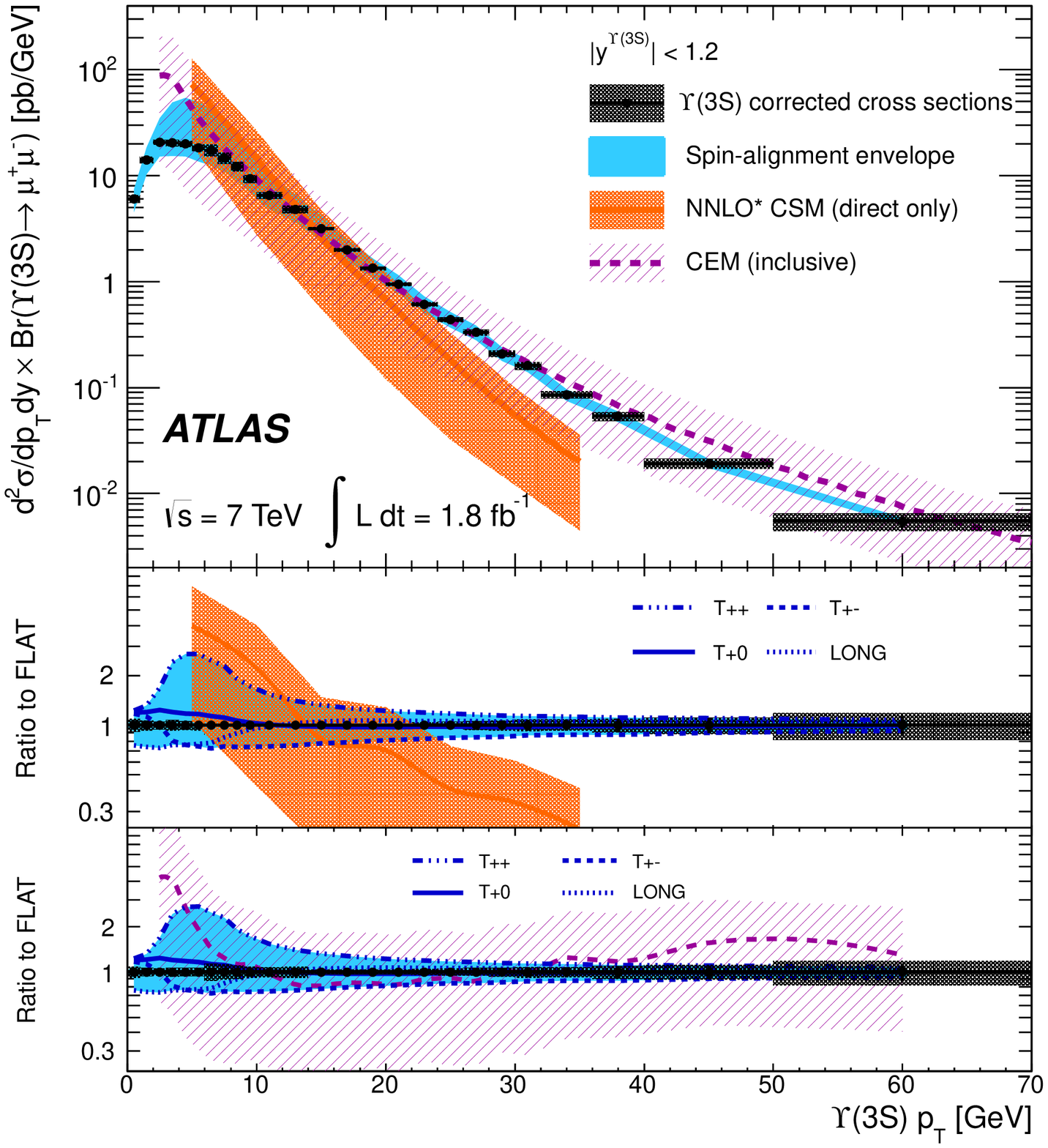}
  \includegraphics[width=0.48\textwidth]{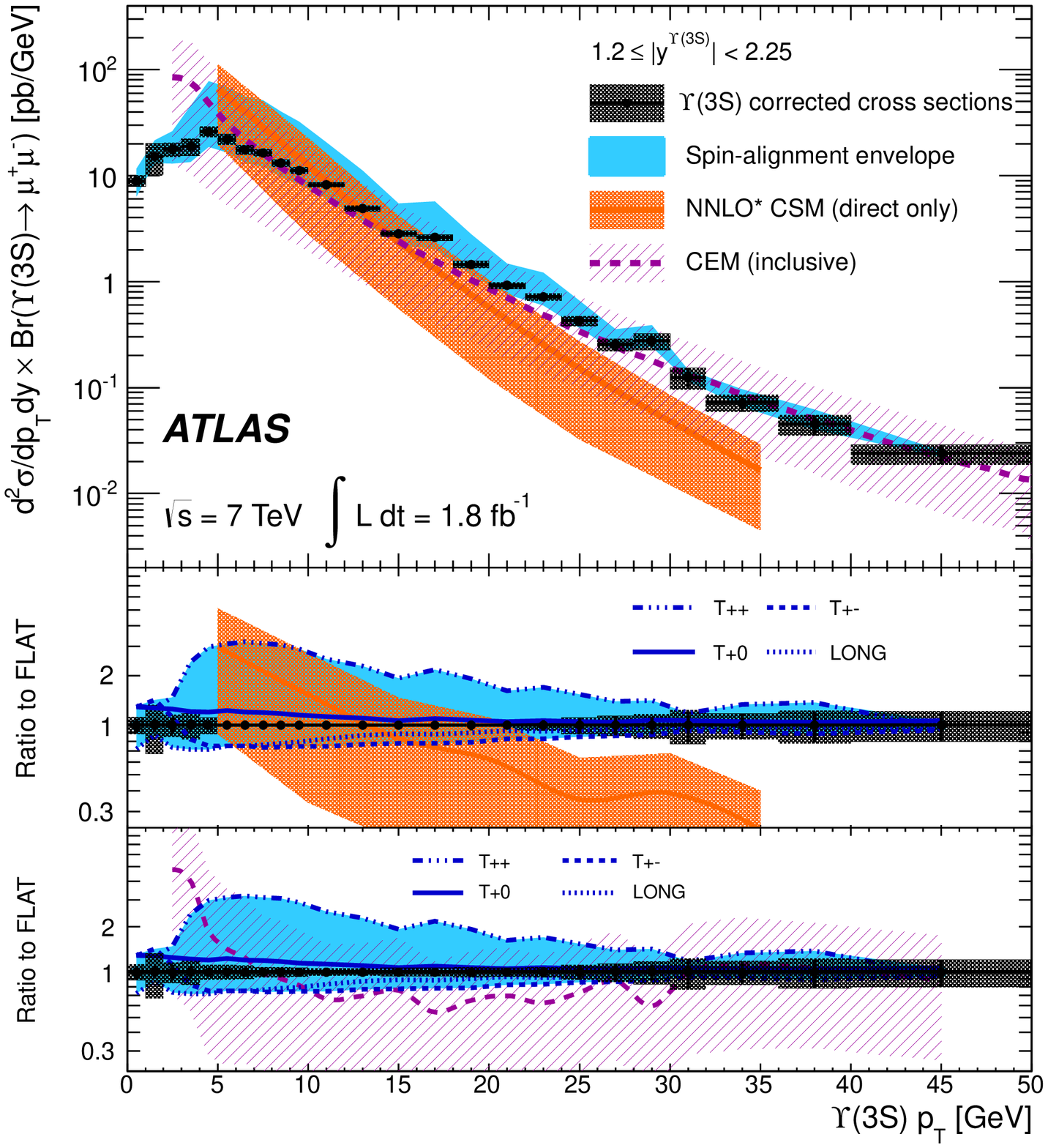}
  \caption{
  As for Fig.~\ref{fig:TotXS-1S}, but for \UpsXS{3} production. 
  }
  \label{fig:TotXS-3S}    
  \end{center}
\end{figure}

\begin{figure}[phtb]
  \begin{center}        
    \includegraphics[width=0.46\textwidth]{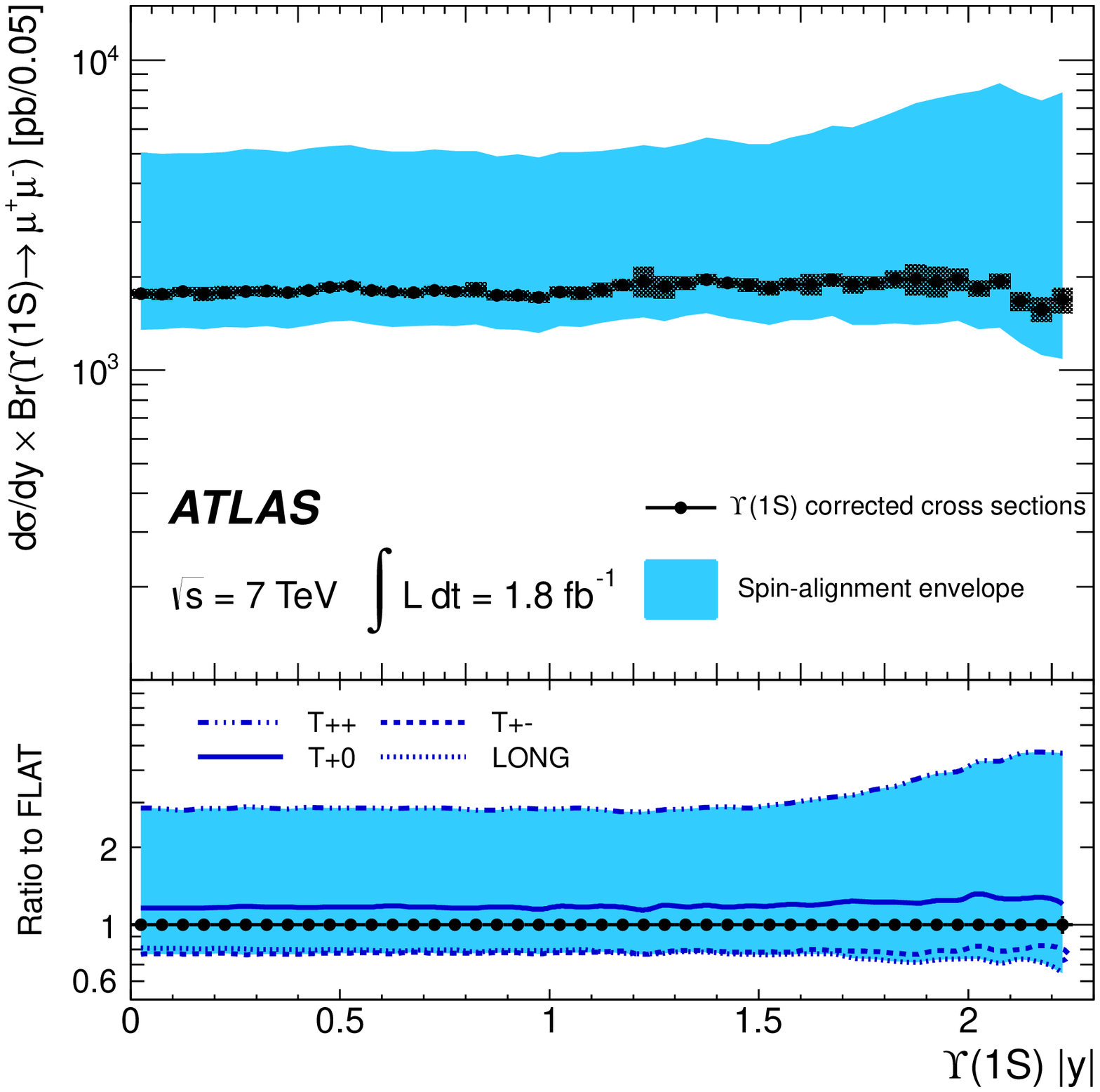}
  \caption{
    Differential cross sections multiplied by the  di-muon branching fraction,
    \XSyB, for \UpsXS{1} production extrapolated to the full phase space.
    Points with error bars indicate results of the measurements with
    total (statistical and systematic) uncertainties. The maximal envelope of variation of the result due to spin-alignment uncertainty
    is indicated by the solid band.
    The variations of the cross-section measurement under the four anisotropic spin-alignment scenarios are shown in the lower pane as ratios to the unpolarized scenario.
  }
  \label{fig:TotXS-rapidity-u1s}    
  \end{center}
\end{figure}

\begin{figure}[phtb]
  \begin{center}        
    \includegraphics[width=0.46\textwidth]{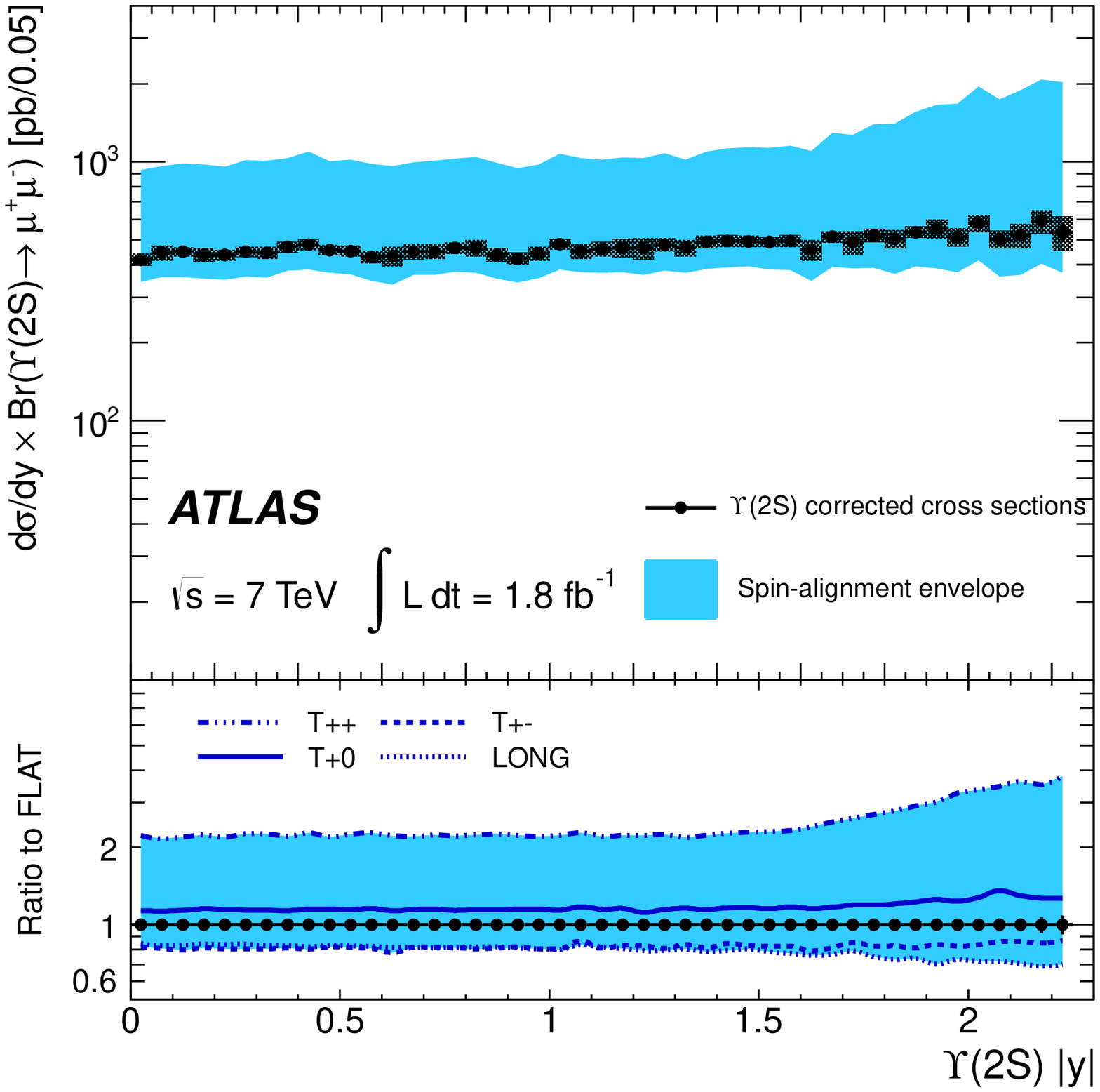}
  \caption{
  As for Fig.~\ref{fig:TotXS-rapidity-u1s}, but for \UpsXS{2} production. 
  }
  \label{fig:TotXS-rapidity-u2s}    
  \end{center}
\end{figure}

\begin{figure}[phtb]
  \begin{center}        
    \includegraphics[width=0.46\textwidth]{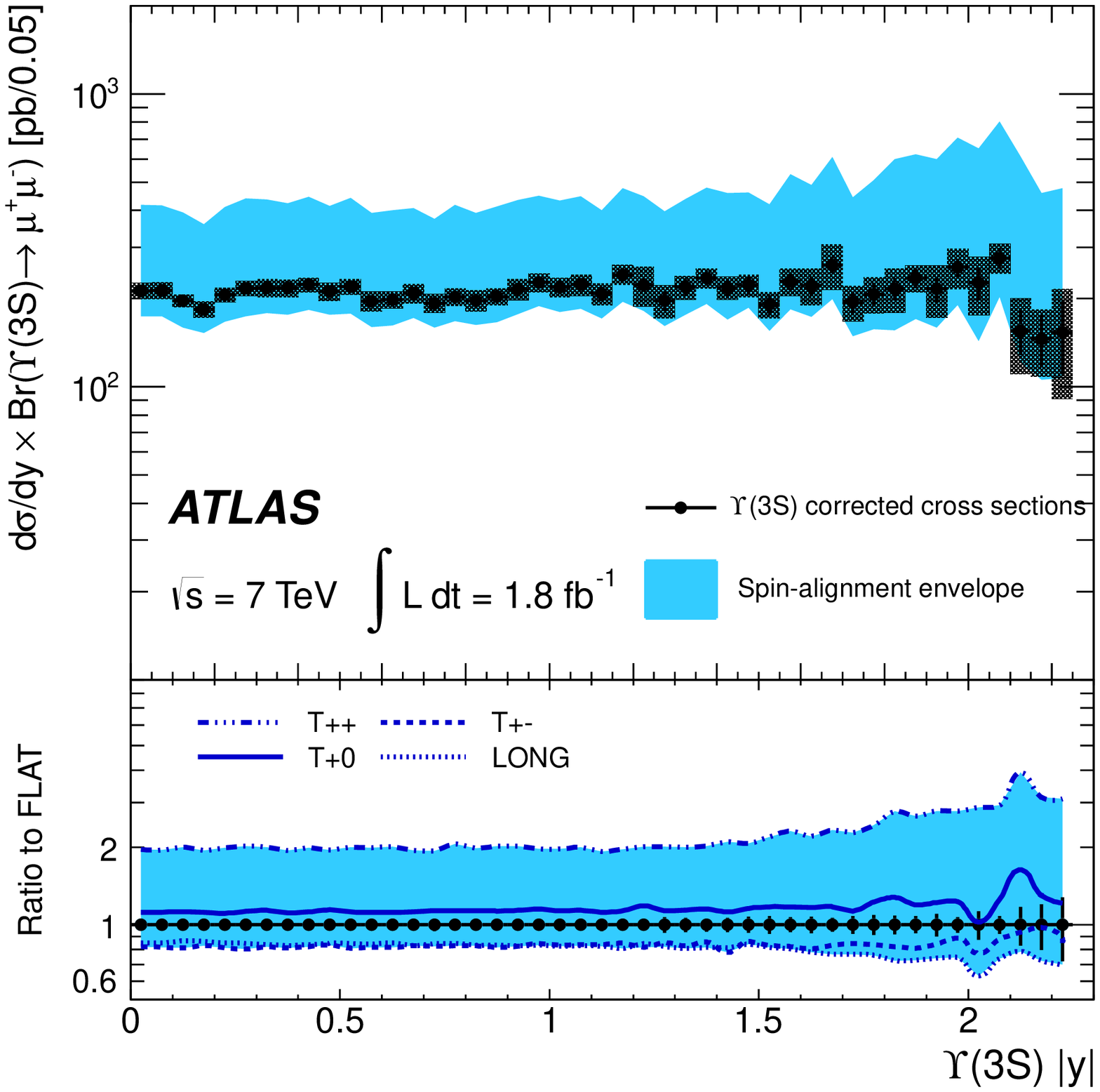}
  \caption{
  As for Fig.~\ref{fig:TotXS-rapidity-u1s}, but for \UpsXS{3} production. 
  }
  \label{fig:TotXS-rapidity-u3s}    
  \end{center}
\end{figure}

Figures~\ref{fig:TotXS-1S}--\ref{fig:TotXS-rapidity-u3s} show the differential cross sections as a function of $p_{\rm T}$ and rapidity
for each of the $\varUpsilon$ states in comparison with theoretical predictions. The effects of varying spin-alignment assumptions from the nominal assumption of
isotropic muon angular distributions independent of $\theta^*$ and $\phi^*$ are indicated by a shaded band.

Clearly, spin-alignment has a large effect on the \Ups\ production cross sections, especially at low \pTUps,
and in particular if there is a non-trivial azimuthal component to the spin-alignment. 
New results\,\cite{CDF:upsilon,CMS:upsilonpol} of $\varUpsilon$ spin-alignment from CDF and CMS
suggest that the spin-alignment is consistent with unpolarized production. 
Our central assumption of isotropic \Ups\ decays is consistent with these results.
Nevertheless, as these spin-alignment measurements are made at different center-of-mass energies or in a restricted phase space in
both \pTUps\ and rapidity with respect to measurements presented here, we provide the results under a variety of polarization scenarios so that the
impact of spin-alignment on the corrected cross sections can be quantified across the full range of study.

The contributions of the five polarization scenarios can be seen in the lower panes of each plot
where the ratio of the differential cross section under these spin-alignment assumptions to the unpolarized scenario is shown. Across the whole $p_{\rm T}$ range studied
the envelope is bounded from above by the $T\mathrm{++}$ ($\lambda_{\theta}=+1, \lambda_{\phi}=+1, \lambda_{\theta \phi}=0$) scenario with a maximal $\phi^*$ variation.
From below, the cross-section envelope is bounded by fully longitudinal spin-alignment at very low $p_{\rm T}$, with the $T\mathrm{+-}$ 
($\lambda_{\theta} = +1, \lambda_{\phi} = -1, \lambda_{\theta \phi} = 0$) scenario resulting in the largest downward variation at $p_{\rm T}\gtrsim 4$\,\GeV.
In this measurement we extend the $p_{\rm T}$ range above 45\,\GeV\ where the maximal possible impact due to the unknown spin-alignment
of $\varUpsilon$ is below $\pm 10\%$. This is significantly smaller than the theoretical uncertainties and is of similar magnitude to current experimental uncertainties.
As such, this region will offer a precision environment to compare future theoretical studies of $\varUpsilon$ production to data.

In Figs.~\ref{fig:TotXS-1S}--\ref{fig:TotXS-3S} a comparison is also made to two theoretical predictions of $\varUpsilon$ production.
The first\,\cite{NNLOUpsilon2}, is a QCD-based calculation using the color-singlet mechanism\,\cite{CSM}, referred to as NNLO* CSM and
presumes that \Ups\ meson production occurs via a color-singlet state and includes full corrections up to next-to-leading order
(NLO), as well as some of the most important next-to-next-to-leading-order (NNLO) terms. This inclusion significantly modifies the prediction.
The partial nature of the higher-order calculation limits the applicability of the calculation
to values above a particular $\varUpsilon$ $p_{\rm T}$ threshold 
and increases the sensitivity of the prediction to the choice of
renormalization and factorization scales.
The second prediction, known as the Color Evaporation Model\,\cite{CEM,CEMcalc}, labeled as CEM, 
is a phenomenological model for inclusive $\varUpsilon$ production based on quark-hadron duality. This model assumes that any heavy $Q\bar{Q}$
pair produced in the initial hard scattering evolves to a quarkonium state if its mass is below the threshold of open heavy-flavor meson
pairs. Predictions of the CEM model involve a single constant that must be determined from corrected cross-section measurements for each
quarkonium state and use a $b$-quark mass of 4.75\,\GeV. As in the case of the CSM, divergences in the predictions restrict the applicability of the model at low $\varUpsilon$ $p_{\rm T}$.
Uncertainties from factorization and renormalization scales are estimated from varying the scales independently up and down by a factor of two and additional uncertainties are estimated
from varying the $b$-quark mass\,\cite{FONLLsyst}.

Color-singlet calculations at NLO/NNLO* predict a largely longitudinal polarization of direct $\varUpsilon$
particularly at high $p_{\rm T}$ (although the effect of feed-down introduces large uncertainties to this prediction).
The Color Evaporation Model offers no explicit prediction of the spin-alignment evolution of $\varUpsilon$ but the nature of the model
suggests no strong polarization should be observed as no single production mechanism dominates.

As can be seen in Figs.~\ref{fig:TotXS-1S}--\ref{fig:TotXS-3S}, the two models provide quite different
descriptions of \Ups\ production. Predictions from CSM are for direct $\varUpsilon$ production only, and so do not account for $\varUpsilon$ production that arises
from feed-down from the production of higher $\varUpsilon$ states or from radiative decays of the $\chi_{bJ}(n\mathrm{P})$. From previous measurements\,\cite{CHIBfeeddown} the
contribution of feed-down to $\varUpsilon(1\mathrm{S})$ production is known to be approximately 50\%, but the $p_{\rm T}$ dependence of the feed-down is not well-known
and cannot be reliably predicted and so no explicit correction is made to the CSM predictions shown.
No correction is needed for the CEM as this is already an inclusive calculation. 
Feed-down contributions are expected to similarly contribute to $\varUpsilon(2\mathrm{S})$ production, but no measurement or
reliable prediction for the relative contribution exists, so we do not apply a correction to the direct $\varUpsilon(2\mathrm{S})$ CSM predictions either. 
For $\varUpsilon(3\mathrm{S})$ there is no feed-down from higher $\varUpsilon$ states. Recently, the ATLAS experiment discovered\,\cite{ATLASCHIB3P} the existence of 
a state or states interpreted as the $\chi_{bJ}(3\mathrm{P})$ below 
the $B\overline{B}$ threshold that are expected to have a significant branching fraction for 
radiative decays into $\varUpsilon(3\mathrm{S})+\gamma$ and thus induce a feed-down contribution to $\varUpsilon(3\mathrm{S})$
(and other $\varUpsilon$ states). As the relative production and decay rates of these states are also as yet unknown no correction is applied to the CSM predictions for 
$\varUpsilon(3\mathrm{S})$ production either.

For each of the three \UpsnS\ states, the NNLO* CSM predictions (considering also the additional normalization uncertainty due
to the poorly-known contributions from feed-down) fit our data well in the moderate \pTUps\ region, but exhibit a 
steeper $p_{\rm T}$ dependence than seen in data. The predictions therefore significantly underestimate the cross section at 
high $p_{\rm T}$. Theoretical developments in the prediction of feed-down contributions may improve this description. 
CEM predictions appear to show a better match with data at high \pTUps.
These predictions underestimate the rate (favouring a smaller choice of renormalization/factorization scale or larger $b$-quark mass) 
and have problems in modeling the shape of the spectrum, particularly at lower $p_{\rm T}$. Shape discrepancies cannot be accommodated within the uncertainties
quoted as changes in the scale choice introduce correlated changes in the prediction as a function of $p_T$.

Figures~\ref{fig:TotXS-rapidity-u1s} to~\ref{fig:TotXS-rapidity-u3s} show the variation 
of the production cross section as a function of absolute $\varUpsilon$ rapidity integrated across all $p_{\rm T}$.
The dependence on rapidity is relatively flat in the interval of rapidities in the ATLAS acceptance. 

Variations between spin-alignment scenarios
lead largely to a change in the normalization of the distributions
with little variation in shape except at high rapidity. There, the $T\mathrm{++}$ scenario leads to an increase in the 
cross section with increasing rapidity, while the fully longitudinal scenario leads to a
drop in cross section at high ($|y|\gtrsim 1.7$) rapidity.

\subsection{Cross section ratios}
\label{sec:ratios}
\begin{figure}[htbp]
\begin{center}
    \includegraphics[width=0.46\textwidth]{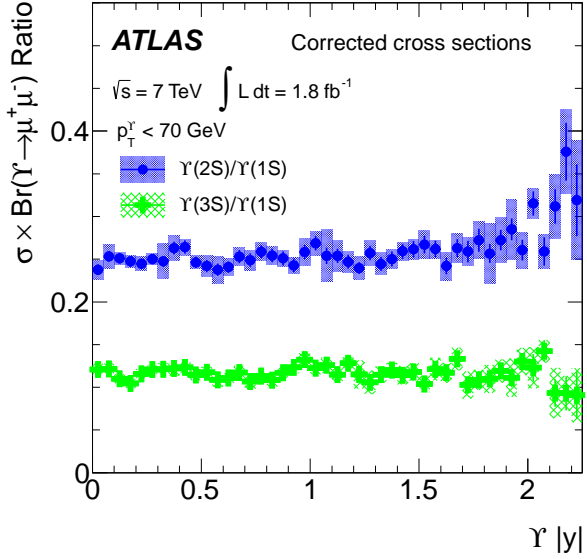}
  \caption{Ratios of differential \UpsXS{2}/\UpsXS{1} and
    \UpsXS{3}/\UpsXS{1} cross sections multiplied by the  di-muon branching
    fractions as a function of di-muon rapidity.
    Points with error bars indicate results of the measurements with
    statistical uncertainties while shaded areas correspond to total uncertainties
    on each point, including systematic effects, but excluding spin-alignment effects.
  }
  \label{fig:Ratio-y}
\end{center}
\end{figure}

\begin{figure}[htbp]
\begin{center}
    \includegraphics[width=0.46\textwidth]{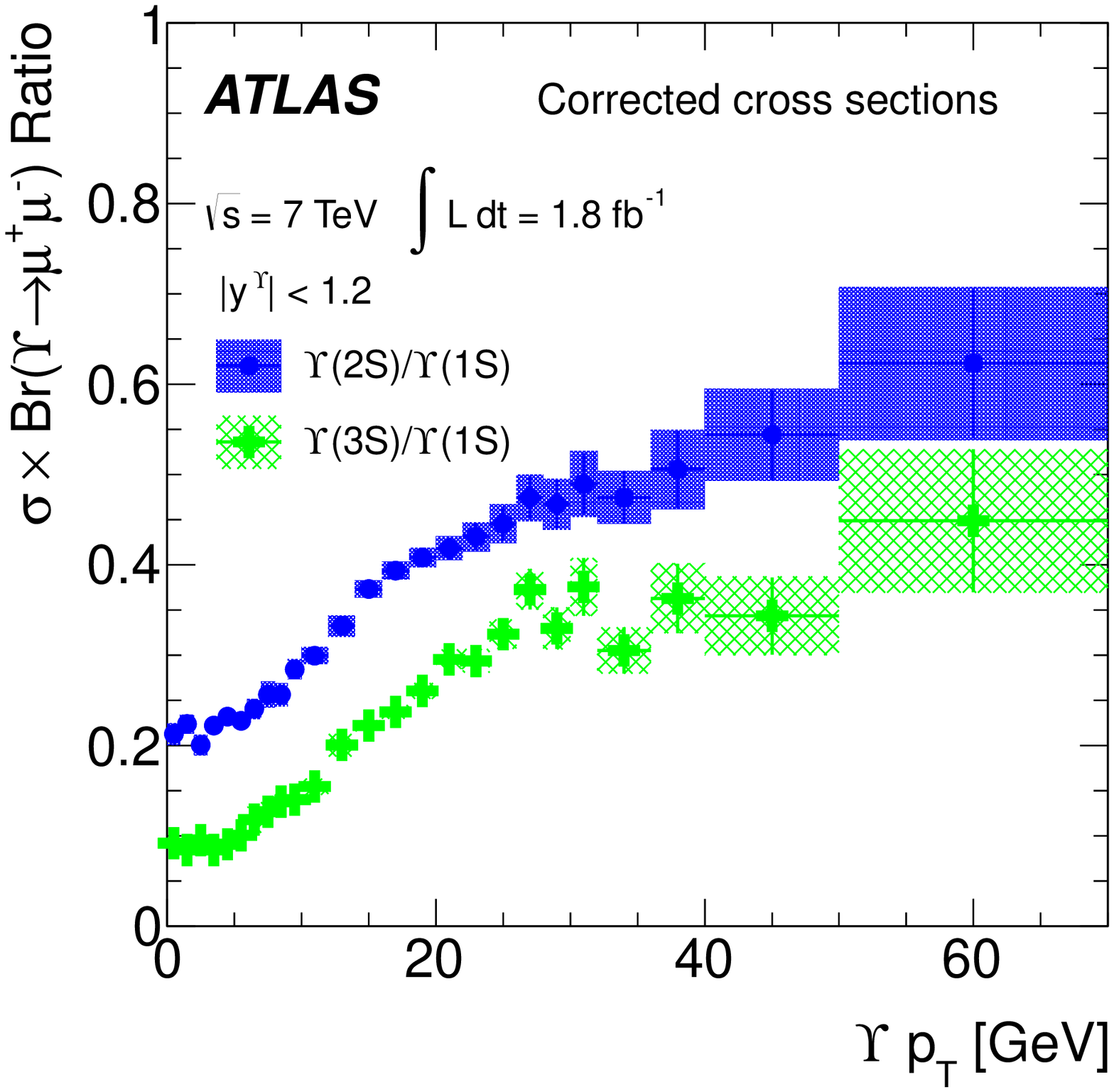}
    \includegraphics[width=0.46\textwidth]{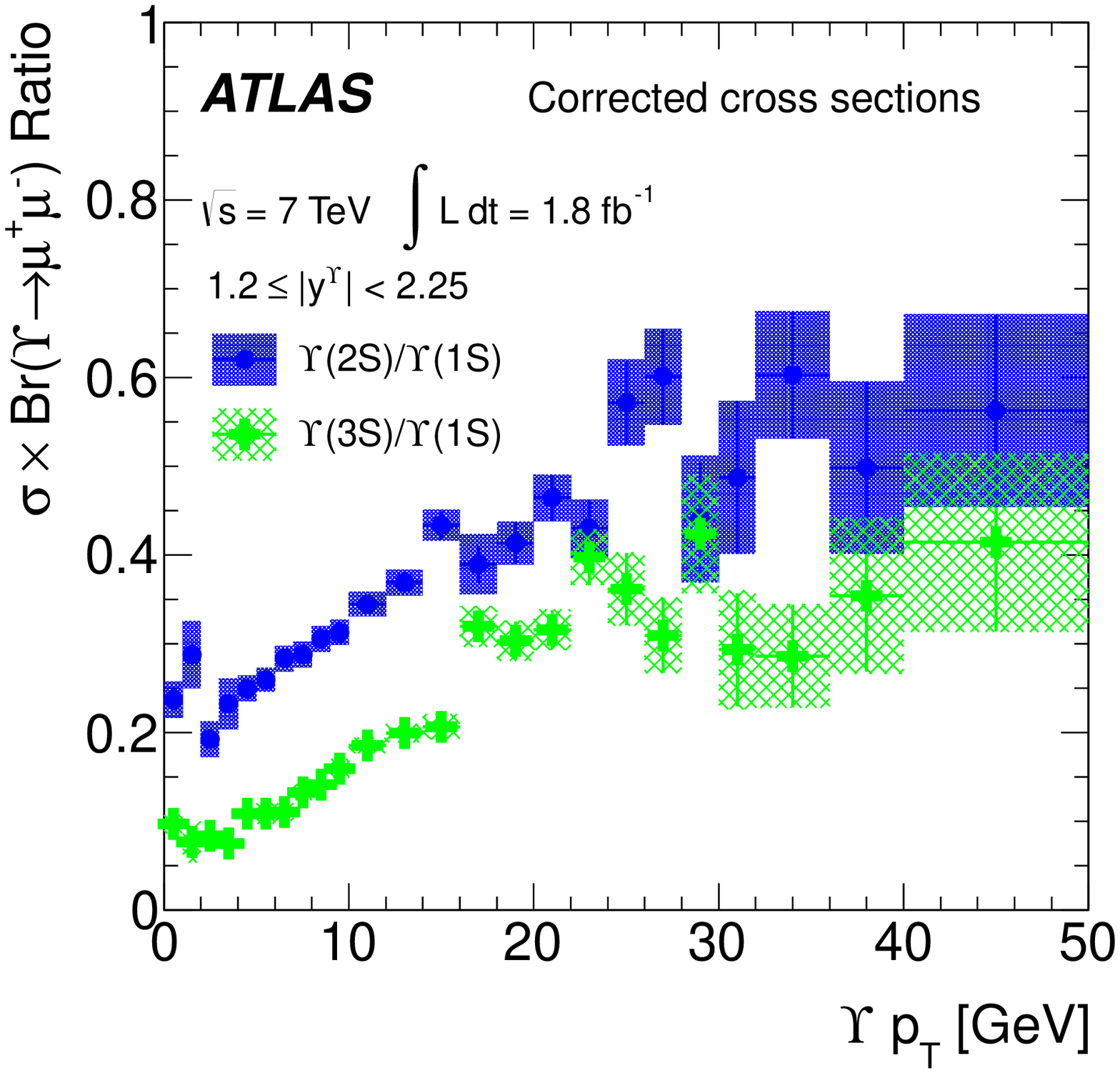}
  \caption{Ratios of differential \UpsXS{2}/\UpsXS{1} and
    \UpsXS{3}/\UpsXS{1} cross sections multiplied by the  di-muon branching
    fractions versus $\varUpsilon$ \pT\ in the (top) central
    and (bottom) forward rapidity regions.
    Points with error bars indicate results of the measurements with
    statistical uncertainties while shaded areas correspond to total uncertainties
    on each point, including systematic effects, but excluding spin-alignment effects.
  }
  \label{fig:Ratio}
\end{center}
\end{figure}

From our differential cross-section measurements, we explore the \pT\ and rapidity dependence of \UpsXS{2} and
\UpsXS{3} production relative to the \UpsXS{1} by deriving the ratios
\begin{align}
\begin{split}
  \frac{d^2 \sigma(pp \to \UpsXS{n}) / d\pT dy}
       {d^2 \sigma(pp \to \UpsXS{1}) / d\pT dy}
  \times
  \frac{\mathrm{Br}({\UpsXS{n} \to \mu^+ \mu^-})}{\mathrm{Br}({\UpsXS{1} \to \mu^+ \mu^-})}
\end{split}
\end{align}
for $\absyUps < 2.25$ with $n$=2,3.
Such observables are sensitive to the magnitude and kinematic dependencies of
feed-down contributions between the three $\varUpsilon$ states.
Results of the differential cross-section ratio measurements are presented in Figs.~\ref{fig:Ratio-y} and~\ref{fig:Ratio}. 

These measurements are made under the assumption of unpolarized
\UpsXS{n} mesons, and take into account statistical correlations between the fitted
numbers of \UpsXS{1}, \UpsXS{2} and \UpsXS{3} mesons.
Systematic uncertainties are estimated by varying acceptance, efficiency, and fit model assumptions coherently in the
numerator and denominator when calculating the ratios, thereby partially canceling uncertainties in the ratio.
Luminosity uncertainties cancel entirely. 

The measured $\varUpsilon(n\mathrm{S})/\varUpsilon(1\mathrm{S})$ ratios are relatively constant in the $0<p_{\rm T}<5$\,\GeV\ interval at $\sim 20\%$ and $\sim 7\%$
respectively for the $\varUpsilon(2\mathrm{S})$ and $\varUpsilon(3\mathrm{S})$. At higher $p_{\rm T}$ a significant and steady rise in the relative
production rates of higher \Ups\ states is apparent, in agreement with measurements by CMS\,\cite{CMS:upsilon}. 
However, at the larger \pTUps\ values (above \pTUps\ of $30$--$40$\,\GeV) accessible for the first time with these measurements, evidence of a saturation in this
rise is apparent, suggesting that we are probing a regime where direct production dominates over contributions from the decays of excited states.
In contrast, the rapidity dependence of these production ratios is quite flat across the full $|y|<2.25$ rapidity interval.

Higher-order color-singlet calculations are not currently able to predict the evolution of these production ratios in \pTUps\
due to the significant feed-down contributions.
At leading order in the quark velocity in the perturbative expansion, the production ratio of the {\em direct}
contributions is proportional to the ratio of the squares of the magnitudes of the wavefunction at the origin (or the partial decay widths), 
multiplied by the  branching fractions to di-muons for each of the states in question. The predicted ratios then are 36\% 
for $\varUpsilon(2\mathrm{S})/\varUpsilon(1\mathrm{S})$ and 29\% for $\varUpsilon(3\mathrm{S})/\varUpsilon(1\mathrm{S})$ for direct production. At high $p_{\rm T}$, 
where the measured production ratio plateaus, the values are somewhat higher than these predictions
suggesting that there is some enhancement (with respect to the simple picture above)
of the $\varUpsilon(2\mathrm{S})$ and $\varUpsilon(3\mathrm{S})$ production relative to $\varUpsilon(1\mathrm{S})$ production.

\section{Conclusions}
\label{sect:Conclusions}
We have measured differential production cross sections and relative
production rates for \UpsXS{1},
\UpsXS{2}, and \UpsXS{3} mesons in \pp\ collisions at \rts\ = 7\,\TeV\ at the LHC
up to $\pTUps<70$\,\GeV\ in the rapidity interval $|y^\varUpsilon|<2.25$.
The possible impact of the \Ups\ spin-alignment on these measured spectra is also quantified.
These measurements are compatible with measurements by the CMS and LHCb collaborations.
We have measured integrated corrected cross sections multiplied by the \UpsMuMu\ branching fractions, within
the rapidity region $\absyUps<2.25$ of \totXSoneS , \totXStwoS , and \totXSthreeS\ for the
\UpsXS{1}, \UpsXS{2}, and \UpsXS{3}, respectively.
Uncertainties correspond to statistical, systematic, and luminosity measurement effects.

These cross-section estimates assume unpolarized production
corresponding to $\lambda_\theta=\lambda_\phi=\lambda_{\theta\phi}=0$ in Eq.~(\ref{eqn:spin}).
If the production polarization is fully transverse or longitudinal with
no dependence on the azimuthal decay angle $\phi^*$, corresponding to
$\lambda_{\theta}=\pm 1, \lambda_\phi=\lambda_{\theta\phi}=0$, the integrated cross sections may vary by up to
$(+19,-23)\%$, $(+18,-21)\%$ and $(+17,-19)\%$ respectively for the $\varUpsilon(1\mathrm{S})$, $\varUpsilon(2\mathrm{S})$ and $\varUpsilon(3\mathrm{S})$. 
However, acceptances corresponding to certain non-zero values of
$\lambda_{\phi}$ can become extremely low, especially at low transverse momenta of the muon pair. This would lead to a significant
increase of the corrected cross sections in these configurations, up to $+217\%$, $+155\%$ and $+126\%$ for the three $\varUpsilon$ states.

We compare our differential cross-section results to predictions from two theoretical approaches describing \Ups\ production. 
Our measurements find both the NNLO* CSM and the CEM predictions have
some problems in describing the normalization and shape of the differential spectra.
In particular, NNLO* dramatically underestimates the rate at high transverse momenta, where the data tend to agree better with the CEM.
The inclusion of $P$-wave feed-down contributions in the theoretical calculation may help to improve the description.
Large scale uncertainties in these predictions allow possible contributions from color-octet terms to contribute to the production rate
in addition to singlet diagrams. The differential production ratios indicate
that the increase in the production of higher \Ups\ states as a function of \pTUps\ relative to the \UpsXS{1} observed previously begins to
saturate at $30$--$40$\,\GeV. Above $\sim40$\,\GeV, the envelope of possible variations in the differential cross sections 
due to spin-alignment is reduced to below $\pm 10\%$. This, along with the expected reduction in feed-down contributions, results in a relatively well-controlled
region in which to study quarkonium production without the dominant experimental and theoretical effects that complicate such studies at lower $p_{\rm T}$.


\section{Acknowledgements}

We thank CERN for the very successful operation of the LHC, as well as the
support staff from our institutions without whom ATLAS could not be
operated efficiently.

We acknowledge the support of ANPCyT, Argentina; YerPhI, Armenia; ARC,
Australia; BMWF and FWF, Austria; ANAS, Azerbaijan; SSTC, Belarus; CNPq and FAPESP,
Brazil; NSERC, NRC and CFI, Canada; CERN; CONICYT, Chile; CAS, MOST and NSFC,
China; COLCIENCIAS, Colombia; MSMT CR, MPO CR and VSC CR, Czech Republic;
DNRF, DNSRC and Lundbeck Foundation, Denmark; EPLANET, ERC and NSRF, European Union;
IN2P3-CNRS, CEA-DSM/IRFU, France; GNSF, Georgia; BMBF, DFG, HGF, MPG and AvH
Foundation, Germany; GSRT and NSRF, Greece; ISF, MINERVA, GIF, DIP and Benoziyo Center,
Israel; INFN, Italy; MEXT and JSPS, Japan; CNRST, Morocco; FOM and NWO,
Netherlands; BRF and RCN, Norway; MNiSW, Poland; GRICES and FCT, Portugal; MERYS
(MECTS), Romania; MES of Russia and ROSATOM, Russian Federation; JINR; MSTD,
Serbia; MSSR, Slovakia; ARRS and MVZT, Slovenia; DST/NRF, South Africa;
MICINN, Spain; SRC and Wallenberg Foundation, Sweden; SER, SNSF and Cantons of
Bern and Geneva, Switzerland; NSC, Taiwan; TAEK, Turkey; STFC, the Royal
Society and Leverhulme Trust, United Kingdom; DOE and NSF, United States of
America.

The crucial computing support from all WLCG partners is acknowledged
gratefully, in particular from CERN and the ATLAS Tier-1 facilities at
TRIUMF (Canada), NDGF (Denmark, Norway, Sweden), CC-IN2P3 (France),
KIT/GridKA (Germany), INFN-CNAF (Italy), NL-T1 (Netherlands), PIC (Spain),
ASGC (Taiwan), RAL (UK) and BNL (USA) and in the Tier-2 facilities
worldwide.

\onecolumngrid
\clearpage
\begin{flushleft}
{\Large The ATLAS Collaboration}

\bigskip

G.~Aad$^{\rm 48}$,
T.~Abajyan$^{\rm 21}$,
B.~Abbott$^{\rm 111}$,
J.~Abdallah$^{\rm 12}$,
S.~Abdel~Khalek$^{\rm 115}$,
A.A.~Abdelalim$^{\rm 49}$,
O.~Abdinov$^{\rm 11}$,
R.~Aben$^{\rm 105}$,
B.~Abi$^{\rm 112}$,
M.~Abolins$^{\rm 88}$
% ...
\bigskip
\\
% ...
$^{11}$ Institute of Physics, Azerbaijan Academy of Sciences, Baku, Azerbaijan\\
$^{12}$ Institut de F{\'\i}sica d'Altes Energies and Departament de F{\'\i}sica de la Universitat Aut{\`o}noma de Barcelona and ICREA, Barcelona, Spain\\
% ...
$^{21}$ Physikalisches Institut, University of Bonn, Bonn, Germany\\
% ...
$^{48}$ Fakult{\"a}t f{\"u}r Mathematik und Physik, Albert-Ludwigs-Universit{\"a}t, Freiburg, Germany\\
$^{49}$ Section de Physique, Universit{\'e} de Gen{\`e}ve, Geneva, Switzerland\\
% ...
$^{88}$ Department of Physics and Astronomy, Michigan State University, East Lansing MI, United States of America\\
% ...
$^{105}$ Nikhef National Institute for Subatomic Physics and University of Amsterdam, Amsterdam, Netherlands\\
% ...
$^{111}$ Homer L. Dodge Department of Physics and Astronomy, University of Oklahoma, Norman OK, United States of America\\
$^{112}$ Department of Physics, Oklahoma State University, Stillwater OK, United States of America\\
% ...
$^{115}$ LAL, Universit{\'e} Paris-Sud and CNRS/IN2P3, Orsay, France\\
% ...
\end{flushleft}



\begin{thebibliography}{99}
\bibitem{ATLASdetectorfootnote}
ATLAS uses a right-handed coordinate system with its origin at the 
nominal interaction point (IP) in the centre of the detector, and the $z$-axis along the beam line. The $x$-axis points from the IP to the centre of the 
LHC ring, and the $y$-axis points upwards. Cylindrical coordinates $(r, \phi)$ are used in the transverse plane, $\phi$ being the azimuthal angle around the beam line. 
Observables labelled ``transverse'' are projected into the $x$--$y$ plane. The pseudorapidity is defined in terms of the polar angle $\theta$ as $\eta = -\ln\tan(\theta/2)$.
Rapidity is defined as $y=\frac{1}{2}\ln\frac{E+p_z}{E-p_z}$ where $E$ is the energy and $p_z$ is the momentum parallel to the beam axis.
\end{thebibliography}
\end{document}